\documentclass[twocolumn]{aastex63}

\usepackage{xcolor}
\usepackage{amsmath}
\usepackage{affils}

\newcommand{\reff}[0]{${\rm R_{eff}}$}
\newcommand{\sersic}[0]{S\'ersic}

\newcommand{\thh}[0]{$^{\rm th}$}

\newcommand{\rrr}[1]{#1}%\textcolor{red}{#1}}

\newcommand{\sbunits}[0]{mag arcsec$^{-2}$}
\newcommand{\muavgmath}[0]{\rm \langle \mu \rangle_{R_{eff}}}%{\bar \mu_{\rm eff,g }}
\newcommand{\muavg}[0]{$\muavgmath{}$}
\newcommand{\hi}[0]{{H}{\small I}}

\newcommand{\mba}[0]{Q_0}
\newcommand{\mca}[0]{S_0}
\newcommand{\refftd}[0]{${\rm R_{eff,3D}}$}
\begin{document}
\shortauthors{Kado-Fong et al.}

    \title{The Intrinsic Shapes of Low Surface Brightness Galaxies: a Discriminant of LSBG Formation Mechanisms}

\author[ 0000-0002-0332-177X]{Erin Kado-Fong}
\affiliation{Department of Astrophysical Sciences, Princeton University, 4 Ivy Lane, Princeton, NJ 08544, USA}
\author{Mihai Petrescu}
\affiliation{Department of Physics, The Ohio State University, 191 W. Woodruff Ave., Columbus OH 43210, USA}
\affiliation{Center for Cosmology and AstroParticle Physics (CCAPP), The Ohio State University, Columbus, OH 43210, USA}
\author{Majid Mohammad}
\affiliation{Department of Physics, Colorado School of Mines,
1523 Illinois Street, Golden, CO 80401}
\author[0000-0003-4970-2874]{Johnny Greco}
\altaffiliation{NSF Astronomy \& Astrophysics Postdoctoral Fellow}
\affiliation{Center for Cosmology and AstroParticle Physics (CCAPP), The Ohio State University, Columbus, OH 43210, USA}
\author{Jenny E. Greene}
\affiliation{Department of Astrophysical Sciences, Princeton University, 4 Ivy Lane, Princeton, NJ 08544, USA}
\author[0000-0002-9798-5111]{Elizabeth A. K. Adams}
\affiliation{ASTRON, the Netherlands Institute for Radio Astronomy, Oude Hoogeveensedijk 4,7991 PD Dwingeloo, The Netherlands}
\affiliation{Kapteyn Astronomical Institute, PO Box 800, 9700AV Groningen, The Netherlands}
\author[0000-0003-1385-7591]{Song Huang}
\affiliation{Department of Astrophysical Sciences, Princeton University, 4 Ivy Lane, Princeton, NJ 08544, USA}
\author[0000-0001-8849-7987]{Lukas Leisman}
\affiliation{Department of Physics and Astronomy, Valparaiso University,  1610 Campus Drive East, Valparaiso, IN 46383, USA}
\author{Ferah Munshi}
\affiliation{Department of Physics and Astronomy, University of Oklahoma, 440 W. Brooks St, Norman, OK 73019, USA}
\author{Dimitrios Tanoglidis}
\affiliation{Department of Astronomy and Astrophysics, University of Chicago, Chicago, IL 60637, USA}
\affiliation{Kavli Institute for Cosmological Physics, University of Chicago, Chicago, IL 60637, USA}
\author{Jordan Van Nest}
\affiliation{Department of Physics and Astronomy, University of Oklahoma, 440 W. Brooks St, Norman, OK 73019, USA}

\correspondingauthor{Erin Kado-Fong} 
\email{kadofong@princeton.edu}
  
  \date{\today}

\begin{abstract}
We use the low surface brightness galaxy (LSBG) samples created from 
the Hyper Suprime-Cam Subaru Strategic Program (HSC-SSP, 781 galaxies), 
 the Dark Energy Survey (DES, 20977 galaxies), and the Legacy Survey
 (selected via \hi{} detection in
 the Arecibo Legacy Fast ALFA Survey, 188 galaxies) to 
infer the intrinsic shape distribution of the low surface brightness galaxy
population. To take into account the effect of the surface brightness cuts
employed when constructing LSBG samples, we
simultaneously model both the projected ellipticity and the apparent surface brightness in our shape inference. 
We find that the LSBG samples are well-characterized
by oblate spheroids, with no significant difference between red and blue 
LSBGs. This inferred shape distribution is in good agreement with similar 
inferences made for ultra-diffuse cluster galaxy samples, indicating that 
environment does not play a key role in determining the intrinsic shape of
low surface brightness galaxies. We also find some evidence that
LSBGs are more thickened than similarly massive high surface
brightness dwarfs. 
We compare our
results to intrinsic shape measures from contemporary cosmological simulations, and find that the observed LSBG intrinsic shapes place considerable constraints on the
 formation path of such galaxies. In particular, LSBG production
 via the migration of star formation to large radii
 produces intrinsic shapes in good agreement 
with our observational findings.
\end{abstract}

\section{Introduction}
Low surface brightness galaxies, or LSBGs, are an observationally defined 
galaxy population characterized by 
a low average surface brightness and large on-sky sizes
\citep[e.g. $\muavgmath{}>24.3$ \sbunits{} and \reff{}$>\rrr{2}.5''$][though specific surface brightness and size cuts vary]{greco2018,tanoglidis2020}; 
ultra-diffuse galaxies (UDGs) are LSBGs with large physical effective radii 
\citep[e.g. \reff{}$>1.5$ kpc,][ -- UDG identification thus requires a distance
measurement]{vandokkum2015}. 
The origin and physical properties
of this extreme tail of the dwarf population is still 
a matter of debate -- 
the formation path of low surface brightness galaxies (LSBGs) and their 
relationship with the general galaxy population has been a topic of sustained 
interest since their discovery \citep{sandage1984, mcgaugh1995, dalcanton1997}.

\begin{figure*}[htb!]
\centering     
\includegraphics[width=\linewidth]{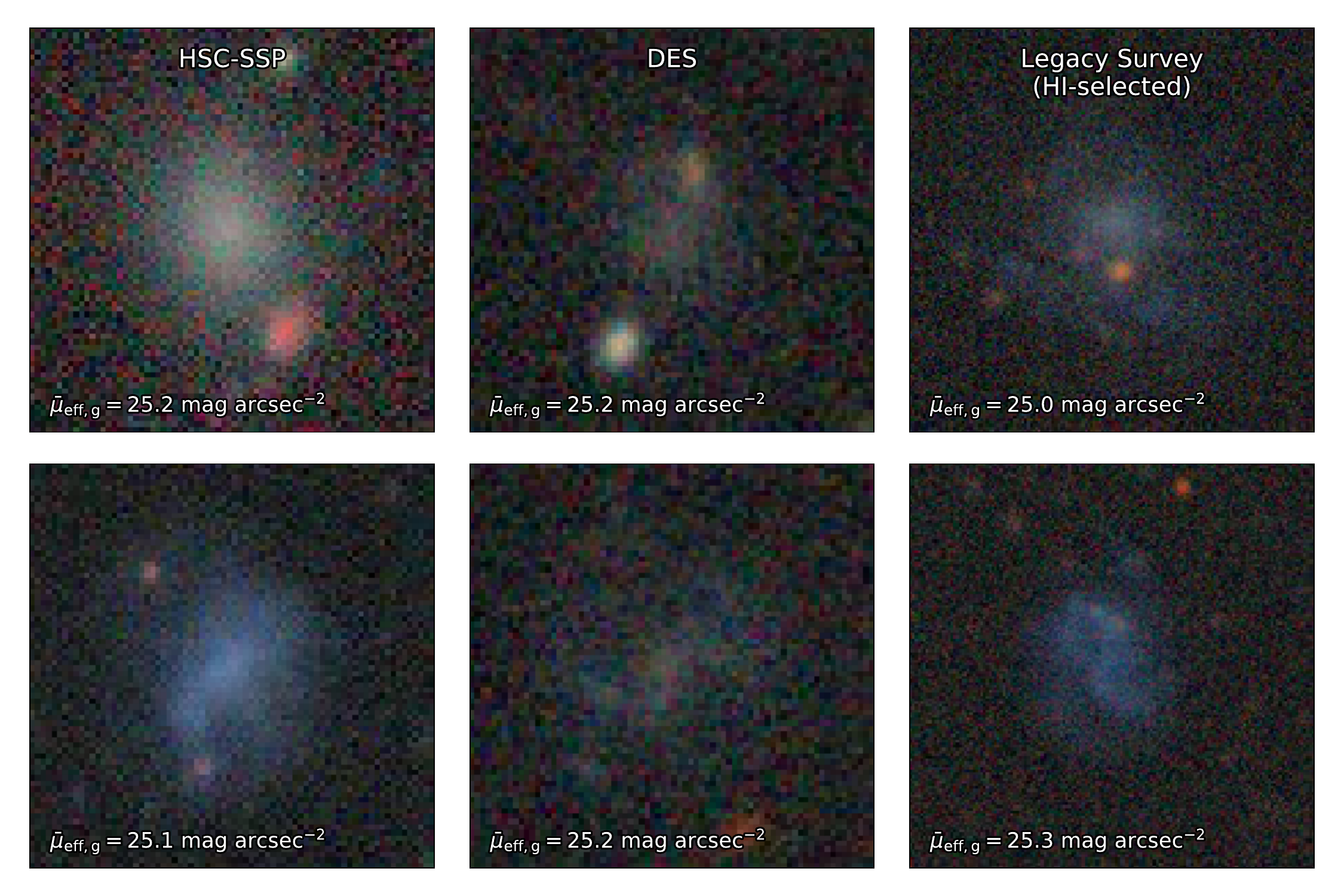} 
%\vspace{-25pt}
\caption{ 
    Example $grz$ RGB cutouts from the three imaging sets considered
    in this work. The top row shows an example from the red end of the
    sample ($(g-r)>\langle (g-r)\rangle_{75}$, where $\langle \rangle_{XX}$ indicates the $XX^{\rm th}$ percentile), while the bottom 
    shows an example from the blue end ($(g-r)<\langle (g-r)\rangle_{25}$).
    Note that because the Legacy Survey UDG sample was originally selected
    to have \hi{}, the red end of the distribution is still relatively blue
    in absolute terms. Each example in the figure is chosen to have
    approximately the same mean effective g-band surface brightness
    of $\muavgmath\sim 25$ \sbunits{}. The cutouts of the HSC and
    DES imaging are 20'' on a side, while the Legacy Survey cutout is
    40'' on a side. \rrr{The galaxies shown for each sample are (top panel, bottom
    panel): HSC (LSBG-325, LSBG-211), DES (ID224789651, ID214498053), and
    Legacy (AGC114754, AGC336397).}
    }
\label{f:images}
\end{figure*}

Present day simulations have proposed
several pathways for the formation of LSBGs and UDGs. 
It has been proposed
that UDGs populate the high-spin tail of dwarf-mass dark matter halos
\citep{amorisco2016,rong2017, liao2019}, are formed via vigorous star
formation feedback and outflows \citep{dicintio2017, chan2018, jiang2019}, are formed via 
effects induced by high-density environments \citep{jiang2019, tremmel2020} or
are the product of early major mergers which trigger the radial
migration of star formation \citep{wright2020}. Given the
significant array of formation scenarios for this class of galaxies, it
is of interest to identify observable quantities which may discriminate
between the proposed formation mechanisms.

The current generation of deep wide-field surveys 
have enabled a new generation of systematic studies of thousands of LSBGs 
over a range of environments \citep{danieli2018, greco2018, tanoglidis2020}.
However, due to the uncertainty in the distances of LSBG samples, 
much of the work on
the inherent physical properties of these low surface brightness galaxies
has been focused on the UDG populations in groups and clusters wherein
the cluster distance may be assumed \citep{vandokkum2015, martinezdelgado2016, vanderburg2016, yagi2016, lee2017, roman2017, roman2017b, danieli2019, mancerapina2019,roman2019, rong2019, zaritsky2019, barbosa2020,prole2021}.
%These studies have shown that UDGs occupy a wide range of halo properties
%\citep{vandokkum2018, vandokkum2019, forbes2021}.
%, with evidence for an
%enhanced globular cluster specific frequency over ``normal''
%dwarfs \citep{vandokkum2016, somalwar2020}.
 
\rrr{The intrinsic, three-dimensional shapes of galaxies provide
key insights into the formation and evolution of galaxy
structure \citep[see, e.g.][]{padilla2008, sanchezjanssen2010, constantin2018, mendezabreu2018, kadofong2020, carlsten2021}}.
\rrr{Indeed, t}he morphology and intrinsic shape distribution \rrr{of LSBGs}
are key properties
that may be explored, even when individual distances are not
known. For ``normal'' high surface brightness galaxies, 
the three-dimensional shape of a galaxy population changes starkly
as a function of both mass and color, producing the 
familiar color-morphology bimodality \citep[see, e.g.][]{padilla2008}. 
In the dwarf mass regime, there is evidence for a transition in
stellar intrinsic shape
from thick disk to oblate spheroid as a function of stellar mass
\citep{sanchezjanssen2010, kadofong2020}. Studies in nearby groups and
clusters suggest that UDGs in high density environments are generally 
oblate spheroids, though the impact of the environment itself is difficult
to ascertain without an analogous sample in less crowded environments.
How the structure of the
general LSBG population relates to that of high surface brightness galaxies 
remains largely unexplored.

\begin{figure}[htb!]
\centering     
\includegraphics[width=\linewidth]{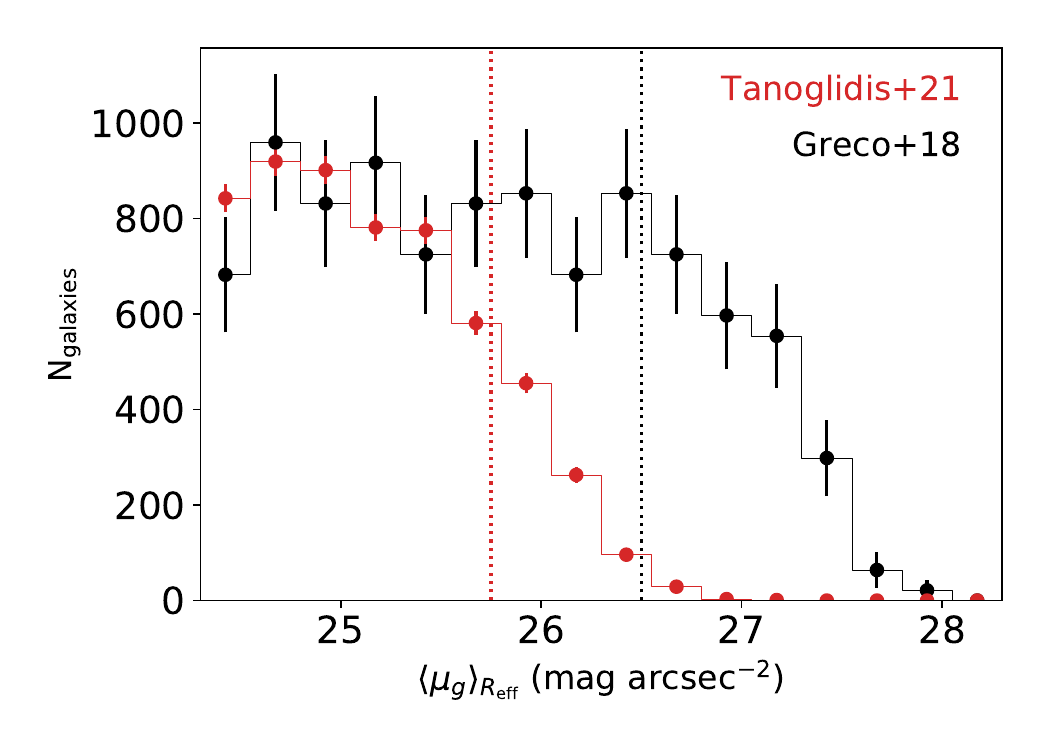} 
%\vspace{-25pt}
\caption{
    Surface brightness distributions of the
    red ($(g-i)>0.64$) galaxies in the DES \citep[][red]{tanoglidis2020} and
    HSC-SSP \citep[][, black]{greco2018} LSBG samples. The $\muavgmath=26.5$ \sbunits{}
    80\% completeness limit for the
    HSC-SSP LSBG sample (Greco et al., in preparation) is shown by the dotted black
    vertical line. We empirically measure the completeness limit of the DES sample as the 
    \muavg{} at which the surface brightness distribution diverges from that of the 
    deeper HSC-SSP
    sample.
    }
\label{f:sblimit}
\end{figure}

\begin{figure*}[htb!]
\centering     
\includegraphics[width=\linewidth]{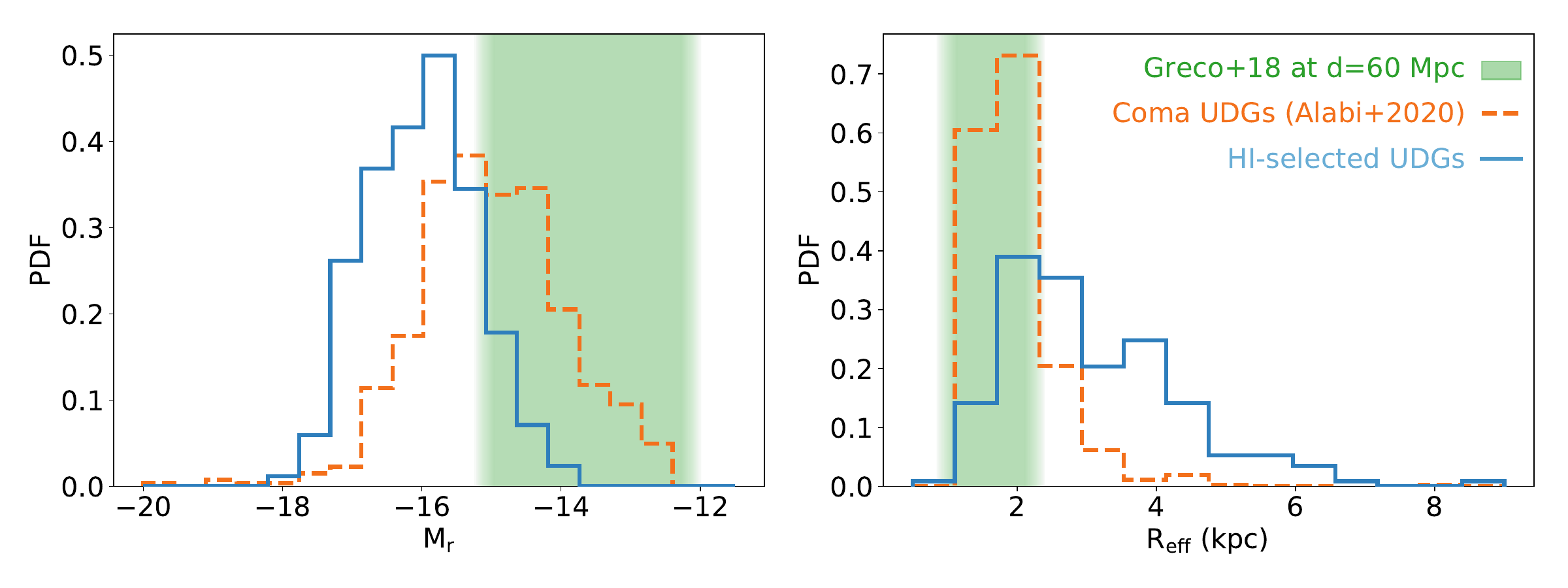} 
%\vspace{-25pt}
\caption{ 
    We compare the distribution in absolute r-band magnitude and 
    physical effective radius for the samples examined in this work.
    We show in green a
    contextualization of the \cite{greco2018} HSC-SSP LSBG sample 
    $\rm M_r$ and effective radius distribution 
    as computed by assuming that all galaxies in the LSBG 
    sample are
    at the cross-correlation median distance of 60 Mpc (Greco et al., 
    in prep.). 
    We show the same quantities computed for the \hi{}-selected
    sample using the \hi{} redshifts measured by 
    \cite{leisman2017} and \cite{janowiecki2019}
    by the unfilled blue histogram. Finally, we show  
    the \rrr{Coma Cluster UDGs of \cite{alabi2020} by
    the unfilled orange histogram (dashed lines).} 
    Though the DES sample is not shown in
    this figure, we expect that the DES sample is similar in mass
    and size to the HSC sample.
    }
\label{f:massfig}
\end{figure*}

In this work we combine three LSBG samples 
to infer the distribution of three-dimensional shapes of the wide-field selected
LSBG population: the sample of
\cite{tanoglidis2020} detected from the Dark Energy Survey (DES), the sample of \cite{greco2018} detected from the Hyper Suprime-Cam Subaru Strategic Program (HSC-SSP),
and the \hi{}-selected sample of \cite{janowiecki2019} selected from the Arecibo Legacy Fast ALFA (ALFALFA) survey and reprocessed using Legacy Survey imaging. 
These three samples provide complementary views of the LSBG population. The DES and HSC-SSP samples are constructed using very similar methods, and while the DES sample includes a larger number of galaxies, the HSC-SSP sample is drawn from deeper imaging data. The \hi{}-selected sample, meanwhile, is selected using a markedly different approach and has associated distances to each object.
In \autoref{s:sample} we detail the samples, as well as the collation and homogenization 
of the DES and HSC-SSP samples, 
In \autoref{s:shapeinference}, we extend the methodology of \cite{kadofong2020} to 
include the effect of intrinsic shape on both ellipticity and surface
brightness and provide an overview of the 
inference machinery used to determine the three-dimensional shape distribution.
We present our main findings for the samples and for color subsets of the
DES sample in \autoref{s:results}, and contextualize our findings with 
previous observational and theoretical work in \autoref{s:discussion}.
Throughout this paper we adopt a standard flat $\Lambda$CDM model 
in which H$_0=70$ km s$^{-1}$ Mpc$^{-1}$ and $\Omega_m=0.3$.

\section{Sample Construction}\label{s:sample}
In this work we use the LSBG catalogs created by \cite{tanoglidis2020} 
and \cite{greco2018}. These catalogs are created using very similar methods,
though the surface brightness limit of the DES imaging is significantly brighter
than the deeper HSC-SSP imaging. We additionally re-analyse the
galaxies of \cite{janowiecki2019} using imaging from the eighth
data release of the Legacy Survey
\citep{dey2019}, both in order to take advantage of the deeper Legacy
Survey imaging and to process the \hi{}-selected sample using a method
consistent with that of \cite{greco2018} and \cite{tanoglidis2020}.
In \autoref{f:images} we show examples cutouts from each imaging set
used in this work. Examples are drawn to span the range of colors
represented by the sample, and are chosen to have approximately the
same mean effective surface brightness. 

\rrr{For all surface brightnesses in this work we refer to the
mean surface brightness within one circularized effective radius, i.e.}
\begin{equation}
\rrr{\rm \langle \mu \rangle_{R_{eff, circ}} = m(<R_{eff, circ}) - 2.5 \log_{10} (\pi R_{eff}^2)},
\end{equation}
\rrr{as derived from single \sersic{} fits, where $\rm R_{eff,circ}$ referes
to the circularized effective radius (${\rm R_{eff,circ}} = a\sqrt{b/a}$, 
where a and b are the observed semi-major and semi-minor axes, respectively).}

\subsection{The HSC-SSP Sample}\label{s:sample:hsc}
The Hyper Suprime-Cam Subaru Strategic Program, hereafter HSC-SSP, is an ongoing widefield 
survey conducted on the 8.2 m Subaru Hyper Suprime-Cam set to cover around 1400 square
degrees to a $5\sigma$ point source depth of $\rm g_{\rm HSC} = 26.6$~mag with a seeing of 0\farcs77
\citep{aihara2019}. The exceptional depth and coverage of this survey make it a powerful
tool for the discovery of low surface brightness structures
\citep[see, e.g.][]{greco2018, huang2018, kadofong2018, wang2019, kadofong2020a}. 
In this work, we use the catalog of 781 LSBGs found by \cite{greco2018} 
in the first $\sim 200$ deg$^2$ of the survey. 

This LSBG sample was constructed 
using a specialized pipeline detailed in \cite{greco2018}; for the 
reader's convenience, we summarize the main points of the method here. The main aim of
the pipeline is to detect contiguous and extended low surface brightness sources. There
are two main contaminants to this goal: low surface brightness structures
associated with bright galaxies (i.e. tidal features or extended stellar halos), and
faint background galaxies. To remove the former, an iterative thresholding is applied
wherein objects with at least 15\% of their
pixels elevated above 28$\sigma$ over the global background are 
discarded from the sample. To remove the latter, each detection is required to contain
at least 100 contiguous pixels. Single component \sersic{} models are then fit to each
remaining source using the software \textsf{imfit}\footnote{https://github.com/perwin/imfit} \citep{erwin2015} and visually 
inspected. \rrr{This visual inspection removes spurious detections due to
low surface brightness contaminants; these are most typically galactic cirrus,
tidal features from massive galaxies, and wings of bright stars.}
Finally, each galaxy is required to have an effective radius of
\reff{}$>2.5$ arcsec, a $\rm g_{\rm HSC}$-band mean effective surface brightness 
(measured within the circularized effective radius) of $24.3<\muavgmath{} < 28.8$ \sbunits{}, and an ellipticity \rrr{($\epsilon = 1 - \rm b/a$, where b/a is the ratio of the 
semi-minor to semi-major axis)} of $\epsilon<0.7$.

\rrr{The selection function of the \cite{greco2018} sample has been explored in detail 
\cite[][chapter 5]{greco2018thesis} via the injection of mock LSBGs 
(characterized by single \sersic{} profiles) into HSC-SSP imaging.
These mock galaxies are processed through the same \textsc{HSCPipe} \citep{bosch2018}
data reduction pipeline
that was employed for the reduction of the imaging used for the \cite{greco2018} analysis.
These tests in particular explore detection efficiency as a function of 
angular size, surface brightness, and \sersic{} index. They find that the 
detection efficiency of the \cite{greco2018} pipeline as a function of surface brightness
is largely independent of angular size, though at surface brightnesses below the 
80\% completeness limit ($\mu_{\rm eff} = 26.5$ \sbunits{}), detection efficiency is
higher for galaxies with larger angular sizes. Similarly, detection efficiency as a function
of angular size is independent of \sersic{} index for galaxies with angular sizes
greater than \reff{}$\gtrsim 4$\arcsec{}. Detection efficiency is higher for low
\sersic{} indices below this angular size. We thus conclude that for the purpose of our work,
which does not focus on galaxies below this 80\% completeness limit, the dependence of 
detection efficiency on size and \sersic{} index should not greatly affect our results. Indeed,
we have performed the analysis detailed in this work using both the full \citep{greco2018}
sample and a surface brightness limited sample ($\muavgmath < 26.5$ \sbunits{}), and find that
our results do not change at statistically significant levels.}

\begin{figure*}[htb!]
\centering     
\includegraphics[width=\linewidth]{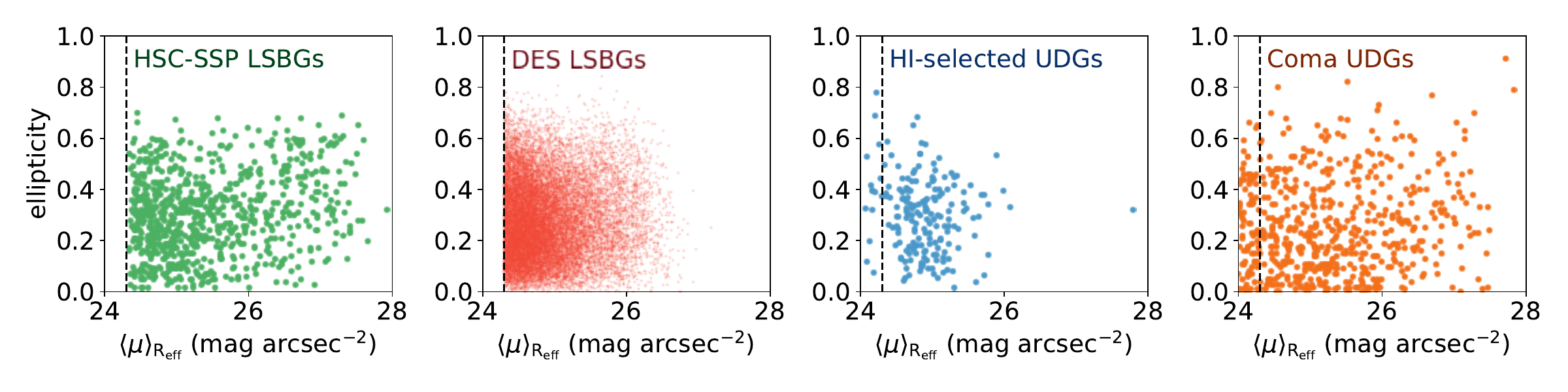} 
%\vspace{-25pt}
\caption{
    \rrr{Ellipticity versus surface brightness for the four samples considered in this work, as 
    labeled. We use these diagnostic plots to demonstrate that the distribution of ellipticity
    does not change over the range of surface brightnesses considered in this work, supporting our
    assumption that the full sample can be characterized by a single distribution in 
    three-dimensional shape, and to visually illustrate the different depths of the
    four samples.}
    }
\label{f:sbell}
\end{figure*}

\subsection{The DES Sample}
Though the depth of HSC-SSP allows for the discovery of very faint LSBGs, the restricted
area of the \cite{greco2018} catalog makes the 3D shape inversion problem intractable for
subsets of the catalog sample. To overcome this limitation, we will also utilize a sample of
20,977 LSBGs selected from
$\sim5000$ deg$^2$ in the first three years
of the Dark Energy Survey (hereafter DES).
The published version of \cite{tanoglidis2020} contains
23,790 LSBGs with a mean effective surface brightness cut of
$\muavgmath{} > 24.2$ \sbunits{}. We use an earlier version of the catalog with a mean effective surface brightness cut of
$\muavgmath{} > 24.3$ \sbunits{}, consistent with \cite{greco2018},
as measured by \textsf{SExtractor}. Both versions of the catalog have been made
publicly available by the DES team\footnote{ \url{https://des.ncsa.illinois.edu/releases/other/y3-lsbg}}.

The sample construction of the \cite{tanoglidis2020} sample is, by design, quite
similar to that of the \cite{greco2018} sample. However, there are a number of 
significant differences, which we will enumerate here for convenience. First, the
DES sample selection was performed on the DES Y3 Gold coadd object catalog 
(v2.2; Sevilla-Noarbe et al. 2020, in preparation). Second,
they employed a support vector machine (SVM) classifier trained from an initial 
visual classification in order to remove contaminant objects. 
Remaining false positives were then removed via visual inspection. 
\rrr{Like the
\cite{greco2018} sample, this visual inspection removes common low surface
brightness contaminants from the final sample.}
Finally, the 
size and average surface brightness cuts were made on the \textsf{SExtractor} measurements,
not on the single component \sersic{} model fits. 

Though a direct measurement of the completeness of this catalog has not yet been made,
we can make an empirical estimate using the surface brightness distribution of
the \cite{greco2018} sample, which is drawn from a deeper survey, as a benchmark.
In particular, the surface brightness distribution of the red ($g-i>0.64$) LSBGs in
the \cite{greco2018} sample is consistent with being flat down to their
80\% completeness limit in surface brightness. We therefore estimate the 80\% completeness
limit of the DES LSBG sample to be at the point where the surface brightness 
distribution of red galaxies in the DES sample diverges significantly from that of the
HSC-SSP sample. \rrr{Because the surface brightness completeness of the HSC sample has 
been extensively tested via mock injections, we base our completeness estimate of the
DES sample on the divergence in surface brightness limit from the HSC sample.}
The red LSBG surface brightness distribution for the two samples are
shown in \autoref{f:sblimit}; using this approach, we find a completeness limit of
$\muavgmath{}\sim25.75$ \sbunits{} \rrr{for the DES LSBG sample}.

Though the initial LSBG sample of \cite{tanoglidis2020} is based on surface brightnesses measured with \textsf{SExtractor}, the final sample selection
is made using surface brightness measurements from \textsf{galfit}. 
%We are using an updated version of the \cite{tanoglidis2020} catalog with 
%surface brightness measurements made in the same way as those of \cite{greco2018},
%for consistency.

\subsection{The ALFALFA/Legacy Sample}
Finally, we include in our analysis
the \hi{}-selected UDG sample of
\cite{janowiecki2019}. The sample is an environment-blind sample originally selected
 UDGs in \hi{} from the ALFALFA survey \citep{giovanelli2005, haynes2011} 
 and measured
the optical properties of the associated galaxies
using imaging from the Sloan Digital Sky Survey
\citep[SDSS,][]{york2000} using the same methods as \cite{leisman2017}, who selected an analogous sample of 
isolated \hi{}-bearing UDGs. Because of the relatively shallow depth of SDSS, \cite{leisman2017} and \cite{janowiecki2019} assumed
an exponential surface brightness
distribution and an ellipticity of $\epsilon=0$. For
this work, we use imaging from the deeper Legacy
Survey \citep[][point source 5$\sigma$ limiting magnitude of $\rm g = 24$]{dey2019} for the 188 galaxies in the 
\cite{janowiecki2019} sample covered by Legacy Survey imaging to  model the surface brightness distributions as single \sersic{} functions
where the \sersic{} index and ellipticity are allowed
to vary. 

Using the \reff{} measurements provided by \cite{janowiecki2019}, we obtained centered cutouts of the galaxies from the Legacy survey's Data Release 8 of size $\rm 20R_{\rm eff}$ on a side. 
%Note that, for some galaxies, the Legacy survey did not have the data necessary to make perfect squares, however this did not affect the quality of the sersic fits.
The \sersic{} fits were carried out using the same methodology as \cite{greco2018}, as summarized in \autoref{s:sample:hsc}, though we used \textsf{sep} \citep{barbary2016} to obtain the initial object segmentations maps. 
Due to the irregularity of the galaxies and large amount of interfering background sources and/or bright star formation knots, ${\sim}40$\% of the object masks were manually adjusted. Measurements from \textsf{sep} were also used to provide initial size and ellipticity guesses to \textsf{imfit}. The new measurements made for this paper will made be public in a forthcoming work (Petrescu, M. et al. in prep.).

\subsection{\rrr{The Coma Cluster Sample}}
\rrr{In order to expand the range of environments probed in this study, we further include
the Coma cluster sample of \cite{alabi2020},
who construct a catalog of Coma Cluster galaxies with a specific focus on
extending the set of known LSBGs in the Coma Cluster. They perform an initial selection using
\textsf{SEXtractor} and refine the sample via visual inspection and structural parameter inference
via \textsf{GALFIT}. We use the catalog measurements of \cite{alabi2020} for this analysis, and
thus caution that the methods used for the structural measurements of this sample
differ significantly than those used for the HSC-SSP, DES, and HI-selected samples, which are
all based on the methodology of \cite{greco2018}. We provide an abbreviated description of the catalog construction method of \cite{alabi2020} for completeness.
First, they perform an initial object detection using 
\textsf{SEXtractor} to remove stars and compact (FWHM$\lesssim 0.5$ kpc) objects. Next, they use a cut in the R-band magnitude-surface brightness plane to identify a sample of galaxies to model. All the remaining galaxies are modelled with one-component \sersic{} profiles using \textsf{GALFIT} \citep{peng2010}. Finally, the 
\sersic{} fits are evaluated, and they remove high
($n>2$) \sersic{} index galaxies that are redder than
$1\sigma$ from the red sequence at the distance of Coma.
We use the structural parameters measured
from a single \sersic{} fit to the Subaru SuprimeCam R band data of \cite{alabi2020}. 
We consider only galaxies in the catalog that are UDGs, with effective radii exceeding 1.5 kpc
and mean effective surface brightnesses fainter than 24 \sbunits{}.}

\subsection{Physical and Observed Properties of the Samples}

As this work centers around the analysis of three
similarly processed but heterogeneously selected samples,
it is informative to compare the physical and observed
properties of the three samples. As the derivation of
many physical properties hinges on a measure of galaxy
distance, we first address existing redshift measurements
for the samples.

A forthcoming analysis of the HSC-SSP catalog estimates a
median source distance of $\sim\!60$ Mpc
(Greco et al., in preparation); we adopt this value in this analysis. 
We show the soft boundaries of the
absolute r-band magnitude and effective radius distributions over the LSBG 
sample if we assume
a fixed distance of 60 Mpc in blue with 
\autoref{f:massfig}. We show the analogous metrics
for the Coma cluster UDG sample of \cite{alabi2020} \rrr{by the orange dashed
histogram}. We stress that the sample range over these physical properties is provided only
to contextualize the general nature of the sample, and should not be interpreted as 
estimates of the absolute magnitude or effective radius of the LSBGs. 
This approach will
not, for example, populate the tails of the absolute magnitude and effective radius
distribution. We note that even if all galaxies were assumed to be at a distance of
100 Mpc, the median absolute r-band magnitude would still be 
$\rm \langle M_r \rangle = -14.8$ ($\rm \langle M_r \rangle = -13.7$
at $\rm d=60$ Mpc).

We do not have a distance estimate for the \cite{tanoglidis2020} sample; because 
the sample selection is modeled after that of \cite{greco2018}, we assume that the
distribution over absolute magnitude and effective radius is similar. Due to the difference in
the surface brightness limits of DES and HSC-SSP (see \autoref{f:sblimit}), this 
assumption is likely incorrect in the details, but should hold as an order of magnitude estimate.

Finally, due to the initial \hi{} selection of the
ALFALFA/Legacy sample, \hi{} redshifts have been measured
for each galaxy; details of these meaurements can 
be found in \citet{leisman2017} and \citet{janowiecki2019}. The r-band absolute magnitude and effective radius distributions for the
\hi{}-selected sample are shown in \autoref{f:massfig}
by unfilled orange histograms. We find that the \hi{}-selected sample
tends to be more luminous and larger than both the HSC and
\cite{yagi2016} samples. 
This is likely because the \hi{} selection was 
sensitive only to the most massive \hi{} disks, as well as because the selection employed
a more stringent physical size requirement than that of \cite{greco2018} (when assuming the median source distance of $\rm d=60$ Mpc). We also note that the \hi{}-selected sample is
significantly more distant than the HSC sample, which is consistent with the \hi{}-selected sample being more luminous.

\rrr{In \autoref{f:sbell}, we show
the distribution of mean effective surface brightness versus ellipticity for the four 
samples considered in this work. These panels show first that there is no global trend in
ellipticity as a function of surface brightness, supporting the assumption that each 
sample can be characterized by a singular unified distribution over three-dimensional shape.
Indeed, the covariance between the surface brightness and ellipticity within our 
surface brightness limits is minimal or consistent with zero for all four samples:
Cov($\langle \mu \rangle_{\rm R_{eff}}, \epsilon)_{\rm HSC}=2.80^{+3.06}_{-3.37}\times 10^{-3}$ \sbunits{},
Cov($\langle \mu \rangle_{\rm R_{eff}}, \epsilon)_{\rm DES}=-1.10^{+0.35}_{-0.34}\times 10^{-3}$ \sbunits{},
Cov($\langle \mu \rangle_{\rm R_{eff}}, \epsilon)_{\rm HI}=-2.74^{+3.65}_{-3.77}\times 10^{-3}$ \sbunits{}, and
Cov($\langle \mu \rangle_{\rm R_{eff}}, \epsilon)_{\rm Coma}=1.32^{+0.70}_{-0.69}\times 10^{-2}$ \sbunits{}.}

\begin{figure*}[htb]
\centering     
\includegraphics[width=.8\linewidth]{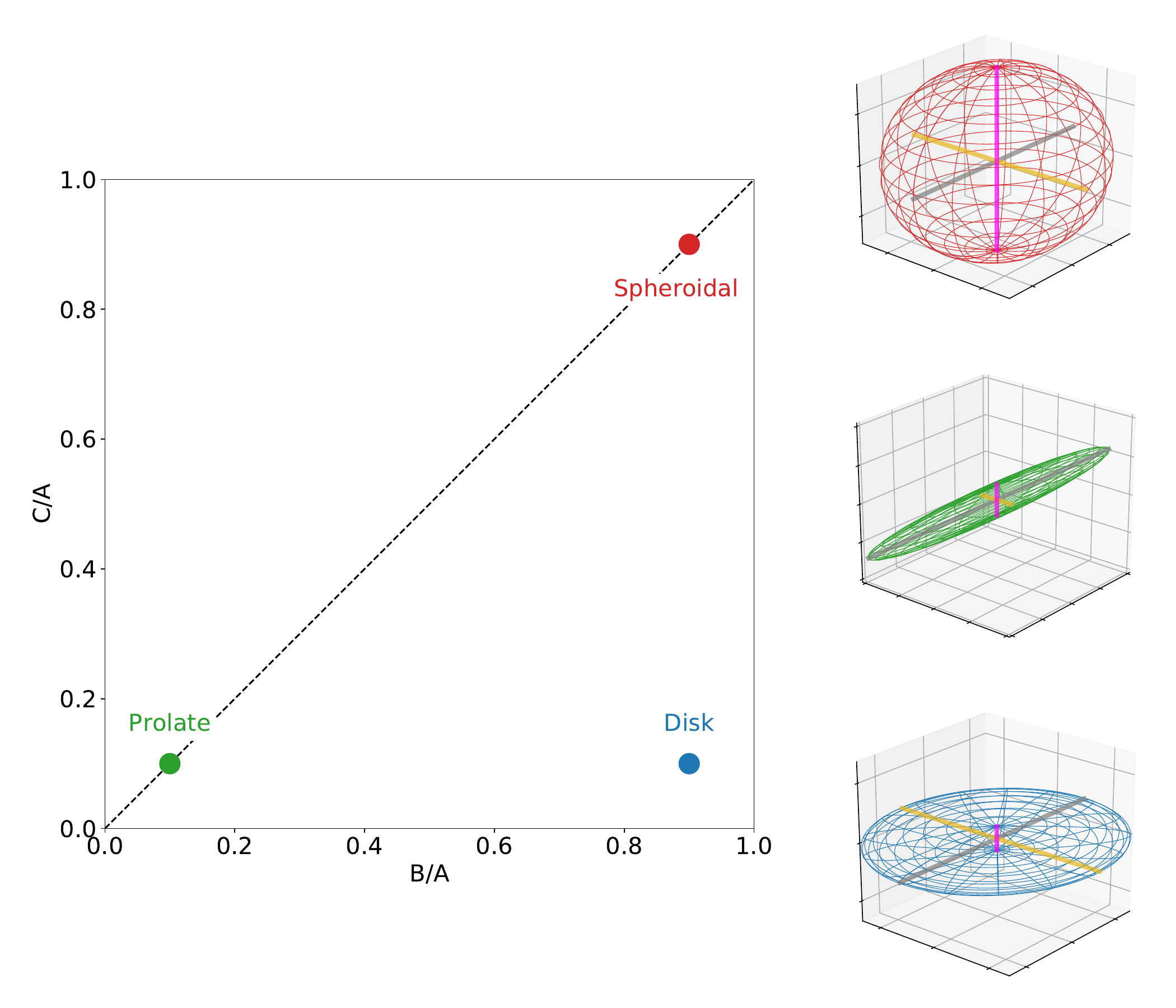}
%\vspace{-100pt}
\caption{
    A schematic diagram to illustrate movement in the $B/A$ vs. $C/A$
    plane. The red, green, and blue points show the position of
    an archetypal spheroidal, prolate, and disky ellipsoid, respectively. The
    axis ratios are ($B/A$,$C/A$) = (0.9,0.9), (0.1,0.1), and
    (0.9, 0.1) for the three cases. 
    At right, we show a three-dimensional
    representation of the ellipsoid that corresponds to each case 
    in the corresponding color.  We additionally show the principal 
    axes A, B, and C as grey, gold, and magenta lines in each panel.
    }
\label{f:schematic}
\end{figure*}

\begin{figure}[htb!]
\centering     
\includegraphics[width=\linewidth]{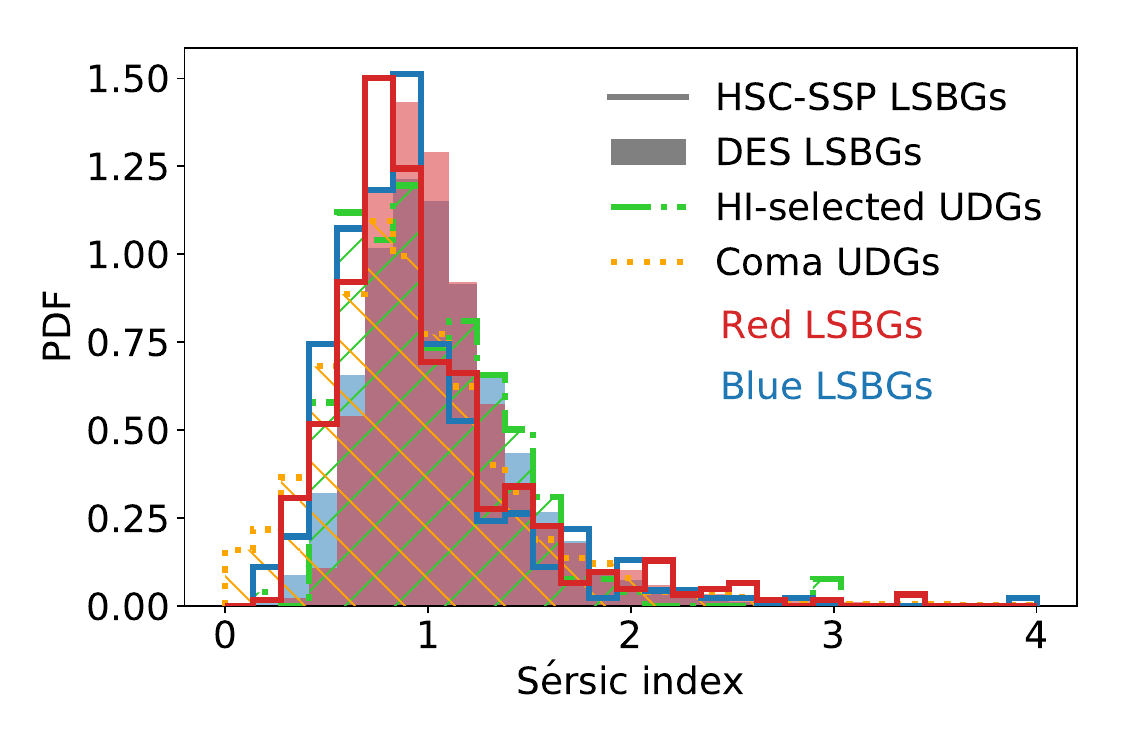}
%\vspace{-25pt}
\caption{ 
    The distribution of \sersic{} indices for the red 
    ($g-i>0.64$) and blue ($g-i<0.64$) HSC LSBGs
    \cite[][filled histograms]{greco2018} and DES LSBGs
    \cite[][unfilled histograms]{tanoglidis2020}. Though
    there is a an offset between the HSC and DES measurements,
    there is no significant difference between the red and 
    blue subsets of each sample. \rrr{We also show the
    HI-selected (dot-dashed green) and Coma (dotted orange)
    UDGs, which are also consistent with the LSBG distribution of
    \sersic{} indices.}
    }
\label{f:sersiccomparison}
\end{figure}

We note that the r band photometry of
the HSC sample is from the HSC r band, the photometry of the \cite{yagi2016}
sample is from the Suprime R band, and the photometry of the \hi{} sample
is from the Legacy Survey r band. 
To gauge the effect of the differences between
these bands, we compute synthetic photometry through
each bandpass for a set of model dwarf spectra taken from dwarfs in
the catalog of \cite{muzzin2013}. We find that the difference between
the bands is minimal, with an average difference between the HSC
and Suprime r bands of 
$(\langle \Delta m \rangle, \sigma_{\Delta m}) = 0.00\pm0.03$ and an
average difference between the HSC and DECam r bands of
$(\langle \Delta m \rangle, \sigma_{\Delta m}) = -0.01\pm0.04$.

\begin{figure*}[htb]
\centering     
\includegraphics[width=\linewidth]{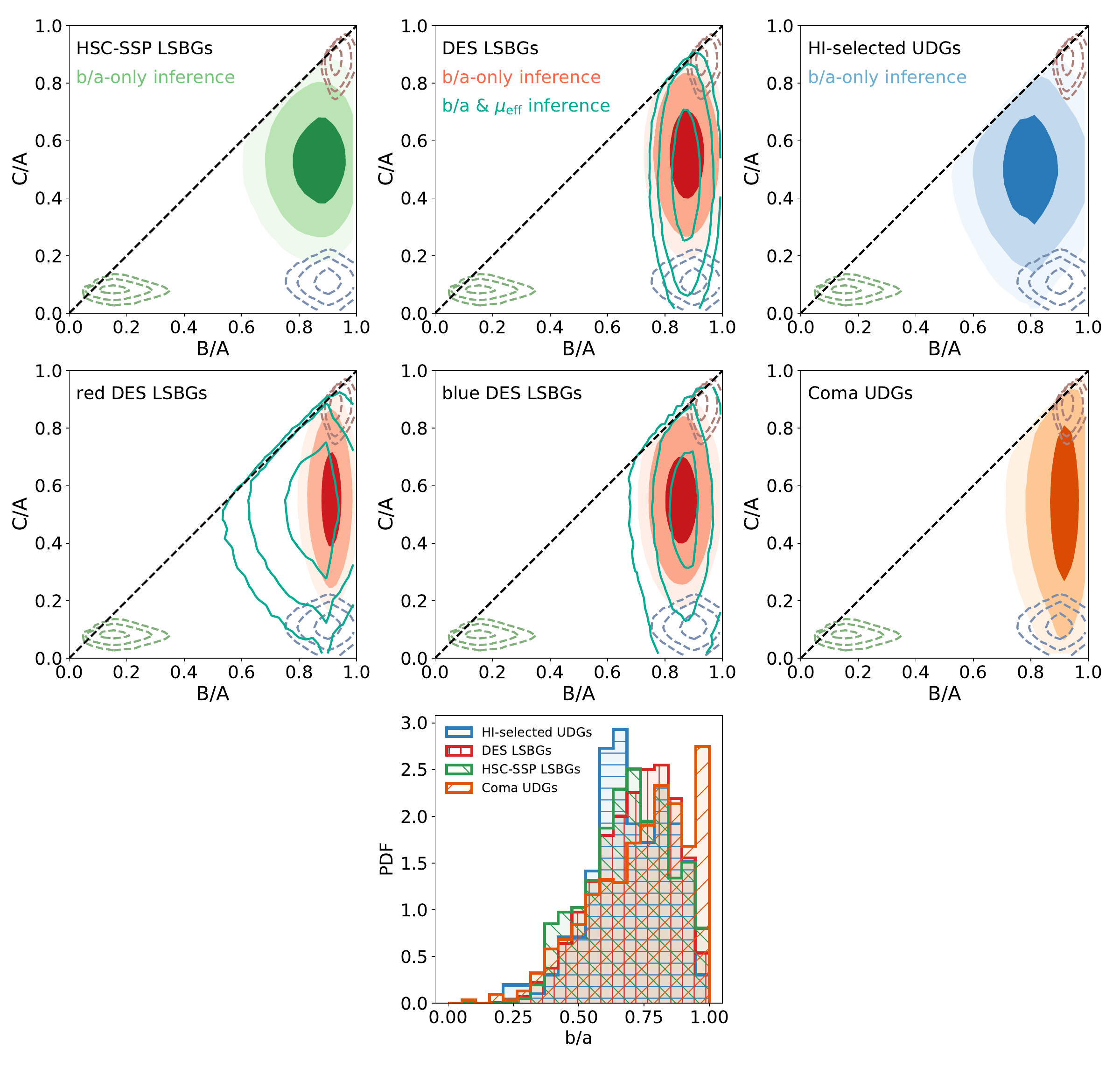}
%\vspace{-25pt}
\caption{ 
    The inferred intrinsic axis ratios for the galaxy samples considered in this work: from the top left,
    the HSC-SSP sample of \cite[green,][]{greco2018}, the DES sample of \cite[red,][]{tanoglidis2020},
    the \hi{}-selected galaxies of \cite[blue,][]{janowiecki2019},
    the red galaxies of the DES sample ($g-i<0.64$),
    the blue galaxies of the DES sample
    ($g-i\geq0.64$), \rrr{and the Coma UDGs of \cite{alabi2020} (orange)}. 
    The filled contours show the 
    inference results when only ellipticity is
    fit, while the unfilled turquoise contours
    in the DES panels show the inference results when both
    surface brightness and ellipticity are fit. 
    Both sets of contours enclose $0.34^2$, $0.68^2$,
    $0.95^2$, and $0.99^2$ of the population
    (corresponding to $0.5\sigma,1\sigma,2\sigma$ and $3\sigma$
    of a multivariate normal). In the panels showing
    the inferred intrinsic axis distributions, the
    unfilled dashed contours show the results of recovery 
    tests for a disky (blue), prolate (green), and spheroidal 
    (red) population via ellipticity-only inference as presented
    in \cite{kadofong2020}. The dashed black line shows the
    definitional $\rm B=A$ boundary.
    The bottom
    right panel shows the projected axis ratio 
    ($\rm b/a=1-\epsilon$) for the three samples used
    in this work. For visual clarity, we do not
    show the projected axis ratio distributions for
    the blue and red subsets of the DES sample here, but
    both are shown in \autoref{f:sbcorrout}.
    Crucially, there 
    is no signficant difference between the inferred shapes of the joint models
    and ellipticity-only models, indicating that incompleteness due to surface brightness limits
    does not induce a significant bias in the observed ellipticity distributions of the LSBGs.
    }
\label{f:inferencecomparison}
\end{figure*}

\begin{figure*}[htb]
\centering     
\includegraphics[width=\linewidth]{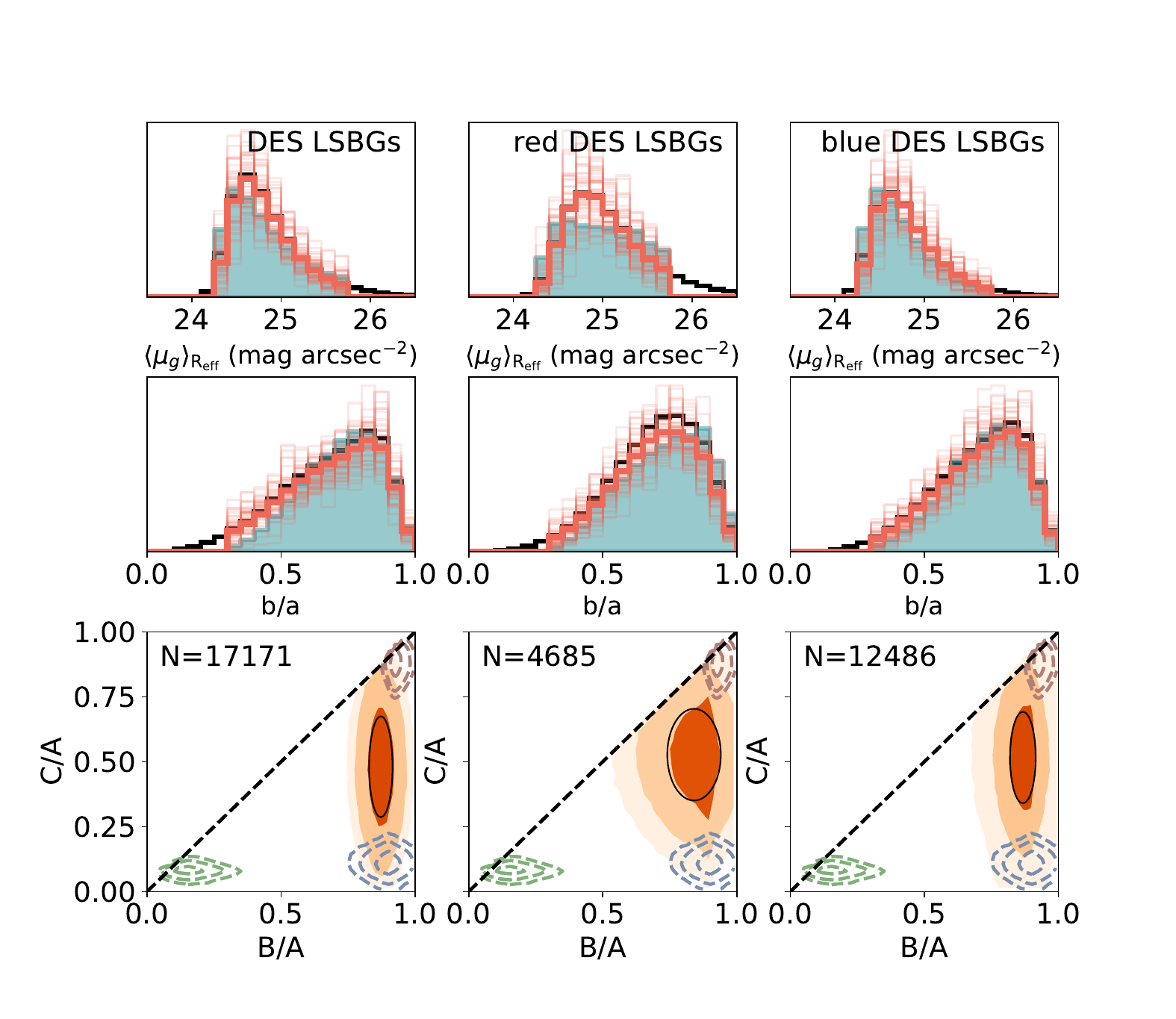}
%\vspace{-25pt}
\caption{ 
    Recovery results for joint inference of the samples; from left to right:
    the full DES sample, red DES LSBGs, and blue
    DES LSBGs. 
    \textit{Top row:} The
    surface brightness distribution for each population. The thick orange curve
    shows our posterior sample (thin orange curves show individual pulls from
    the posterior). The solid teal histograms show the observed population.
    \rrr{Unfilled black} histograms show model projections which fall outside of the
    sample surface brightness limits.
    \textit{Middle row:} the same for the ellipticity distributions. 
    \textit{Bottom row:} we show the intrinsic axis ratio distribution of
    the posterior sample as filled orange contours. The maximum a posteriori
    results are shown by black ellipses. We note that the joint inference
    does not converge for the HSC-SSP and \hi{}-selected samples due to the low number of
    objects; we thus use only the ellipticity-only inference for these samples.
    }
\label{f:sbcorrout}
\end{figure*}

\begin{figure*}[htb]
\centering     
\includegraphics[width=.9\linewidth]{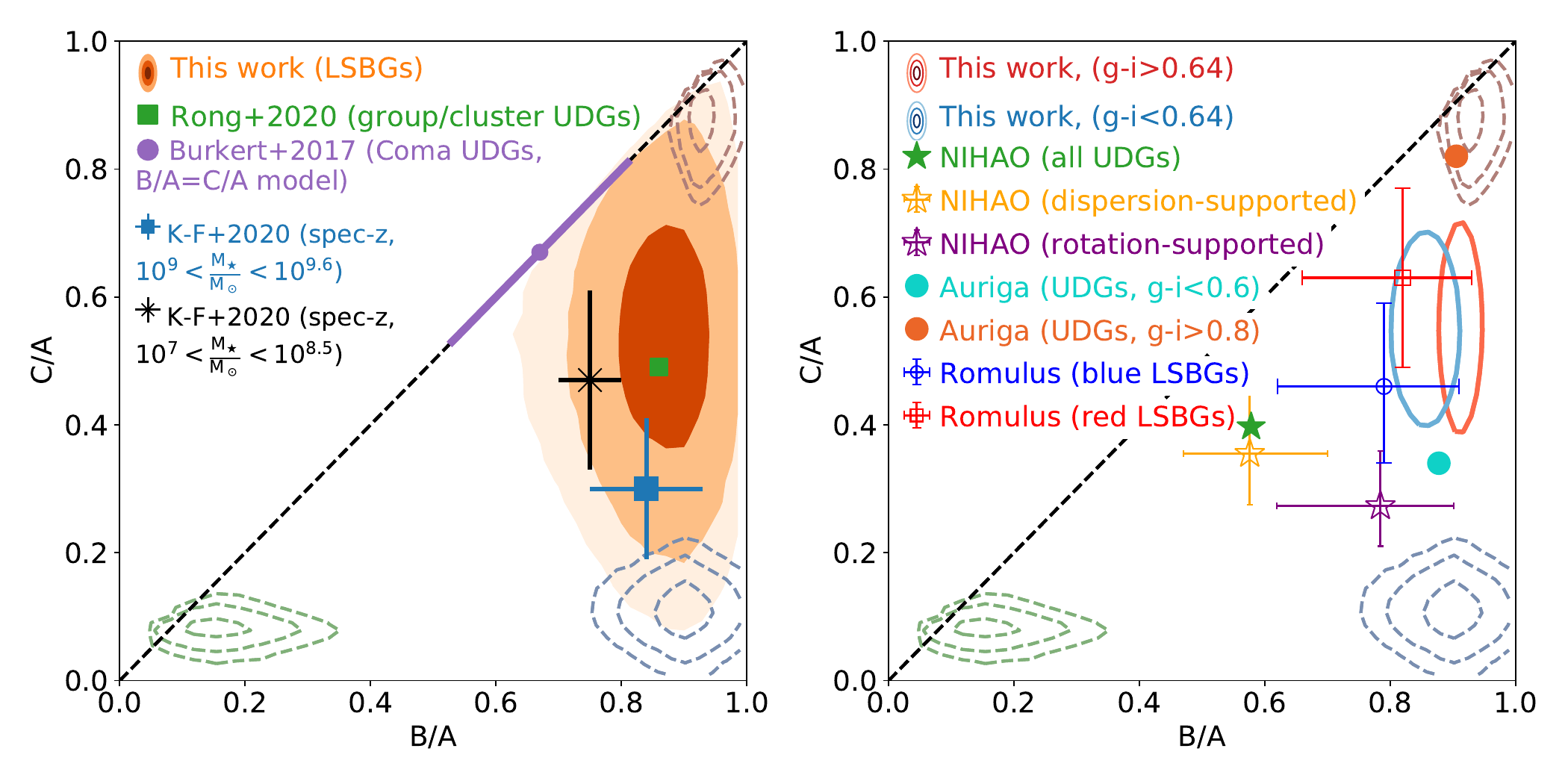} 
%\vspace{-25pt}
\caption{ 
    \textit{Left:} A comparison between the results presented in this work
    (filled orange contours) and observational results from the literature.
    Our results are in good agreement with the group/cluster UDG samples of
    \cite{burkert2017} (who assumes a non-triaxial model where B/A=C/A) and 
    \cite{rong2019}, indicating that the intrinsic shape distribution of LSBGs 
    is not strongly affected by their environment. We also compare our results
    to those of the ``normal'' high surface brightness dwarf populations at 1\reff{} 
    presented in \cite{kadofong2020}. We find that the LSBG sample is significantly
    rounder (higher C/A) than the massive dwarfs of \cite{kadofong2020}. Errorbars on the HSB points show the inferred 1$\sigma$ 
    dispersion of the HSB sample, not the error on the mean values. \textit{Right:} A comparison between the 
    inferred intrinsic shapes of red and blue LSBGs (unfilled red and blue 
    contours, respectively) presented in this work and theoretical predictions
    from the literature. \rrr{For visual clarity, we show only the contours that
    contain 68\% of our inferred LSBG shapes in order to be consistent to the plotted errorbars
    of the Romulus and NIHAO simulations, for which the errorbars span the 16\thh{} to 84\thh{} 
    percentiles.}
    We find that the intrinsic shape distributions
    vary significantly between the simulations considered. In particular,
    Auriga \citep{liao2019} predicts a strong shape divergence between blue
    and red UDGs, while NIHAO \citep{jiang2019, cardonabarrero2020} 
    predicts significantly more
    triaxial \rrr{(for dispersion-dominated UDGs) or flatter (for rotation-supported UDGs)}
    shapes than those inferred from observed LSBGs. Romulus
    appears to well-produce the shape distribution of blue LSBGs, but
    overpredicts the shape evolution between blue and red LSBGs.
    }
\label{f:comparison}
\end{figure*}

\section{Intrinsic Shape Inference}\label{s:shapeinference}
The distribution of intrinsic shapes may be inferred from the observed ellipticity
distribution of a galaxy population by assuming that the galaxies can, at fixed 
radius, be described by ellipsoids with semi-principal axis diameters
A, B, and C where $\rm C\leq B \leq A$. In \autoref{f:schematic}, we show
the positions in B/A-C/A space and 3D renderings of three archetypical 
examples: a disk (blue), prolate (green), and spheroid (red).
We also assume that the LSBG samples
can be described by a single multivariate normal distribution over the principal 
axis ratios B/A and C/A. This method is detailed in \cite{kadofong2020}; we 
summarize the salient points below.

The projected axis ratio, $q=\rm b/a$ where b and a are respectively the 
semi-minor and semi-major axes of the projected ellipse,
for a given ellipsoid is determined solely by the
observer's viewing angle, $(\theta, \phi)$, i.e. 
$\rm q = \mathcal{F}(B/A,C/A,\theta,\phi)$.
The analytic expression for $\mathcal{F}$ was presented by \citet{simonneau1998},
and is reproduced below. First, $(ab)^2$ and $(a^2 + b^2)$ can be rewritten 
as follows:
\begin{equation}
\begin{split}
     a^2b^2=f^2&=(C\sin\theta\cos\phi)^2 +(BC\sin\theta\sin\phi)^2 +\\
     &(B\cos\theta)^2,\\
\end{split}    
\end{equation}

\begin{equation}
\begin{split}
    a^2+b^2=g&=\cos^2\phi + \cos^2\theta\sin^2\phi +\\
    & B^2(\sin^2\phi + \cos^2\theta\cos^2\phi) + (C\sin\theta)^2\\    
\end{split}    
\end{equation}
We now define the quantity $h$ to be
\begin{equation}
    h\equiv \sqrt{\frac{g-2f}{g+2f}},
\end{equation}
such that it may be shown that
\begin{equation}
    \frac{b}{a} = \frac{1-h}{1+h}
\end{equation}
Because the distribution of 
viewing angles is known to be isotropic on the surface of the sphere, 
we can predict the 
projected distribution of $q$ given a choice of intrinsic
shape distribution characterized by $\vec \alpha$ 
by sampling $\phi$ and $\theta$ as follows:

\begin{equation}
\begin{split}
    \phi &\sim \mathcal{U}[0,2\pi]\\
    \nu &\sim \mathcal{U}[0,1]\\
    \theta&=\cos^{-1}\left(2\nu - 1\right)\\
\end{split}
\end{equation}

However, we must also consider the bias imposed by surface brightness selection
inherent in the selection method of LSBG samples. 
To create a low surface brightness galaxy sample, 
one must necessarily make a cut in
surface brightness. Because galaxies are 
three-dimensional objects, it is possible that some galaxies will
be considered ``low surface brightness'' only at certain
viewing angles. This effect is maximized for disky ($\rm C<B\sim A$) 
galaxies, wherein the observed surface
brightness will be systematically lower for face-on views than for edge-on views.
An attempt to invert the 3D shape of such a sample without accounting for this potential
incompleteness could then result in severely biased results.

If one assumes that the LSBGs are well-described by a \sersic{} profile
with $n=1$, the observed
surface brightness of an object as a function of
viewing angle is exactly prescribed by the intrinsic shape of the object and its 
intrinsic stellar density. We find that this assumption is well-supported by the data (see \autoref{f:sersiccomparison}), and adopt a fixed profile with $n=1$ for this work.
We neglect the effect of dust on the observed shape in this framework -- due to the low masses, large effective radii, and low surface densities of the objects in question, it is unlikely that dust will significantly impact the observed shape at 1\reff{}. With these assumptions in hand, by including additional parameters to describe the
distribution of intrinsic density of the LSBGs, we can take into account this
potential surface brightness incompleteness when inferring the intrinsic
3D shapes of the LSBG samples.

To do so, we compute
the projected surface brightness within \reff{} for a given choice of
shape (B/A,C/A), viewing angle ($\theta,\phi$) and intrinsic density $\rho_0$. We find that using
three-dimensional \refftd{} (as opposed to the projected \reff{}) does not strongly affect our results, and considerably
reduces computation time, so choose to use the three-dimensional \refftd{} in this calculation.
We first compute the three-dimensional density using the analytic approximation for the
three-dimensional \sersic{} density profile of \cite{prugniel1997}:
\begin{equation}
        \rho(\frac{r}{\rm R_{\rm eff,3D}}) = \rho_0 (\frac{r}{\rm R_{\rm eff,3D}})^{-p_n}\exp[{-b_n (\frac{r}{\rm R_{\rm eff,3D}})^{1/n}]}
\end{equation}
where 
\begin{equation}
    p_n = 1 - \frac{0.549}{n} +\frac{0.055}{n^2}\\
\end{equation}
and $\Gamma(2n) = 2\gamma(2n,b_n)$. We rotate this grid to align with the line of sight defined
by $(\theta,\phi)$ using the rotation matrix
\[ \mathbb{A} = \begin{bmatrix}
\cos\phi & \cos\theta\sin\phi & \sin\theta\sin\phi \\
-\sin\phi & \cos\theta\cos\phi & \sin\theta\cos\phi \\
0 & -\sin\theta & \cos\theta \end{bmatrix} \]
such that the rotated coordinates $\vec r_r = \mathbb{A} \vec r$ align with the
line of sight along the $\vec z$ axis. We then measure the mean 
surface brightness within 1\reff{}$_{,3D}$. To reduce computation time, we pre-compute the surface brightness as a function of shape (B/A, C/A) and viewing angle 
($\theta,\phi$), sampling uniformly over parameter space in a grid of size $30^4$.
We then linearly interpolate over this grid during inference; we find that our
results are not significantly affected by the adoption of this approximation 
scheme.

In order to jointly fit the ellipticity and mean surface brightness distribution
of our galaxy samples, we modify the Poisson likelihood used in
\cite{kadofong2020} to include the surface brightness distribution:
\begin{equation}
\begin{split}
    \ln{}p({\vec q_{\rm obs}},{ \langle {\vec \mu} \rangle_{\rm R_{eff}, obs}}|\vec\alpha) &= \sum_i n_i \ln{m_i} - m_i - \ln{n_i!}\\
    \vec\alpha &= (\rm \mba{}, \mca{},\mu_{\rho_0}, \sigma_{\mba{}},\sigma_{\mca{}},\sigma_{\rho_0})
\end{split}
\end{equation}
\rrr{where $\rm Q=B/A$ and $\rm S=C/A$.
In our notation, $\rm \mba{}$ and $\sigma_{\mba{}}$
($\rm \mca{}$ and $\sigma_{\mca{}}$)
correspond to the mean and standard deviation of the axis ratio B/A (C/A) assuming a bivariate Gaussian as the functional form for the underlying intrinsic shape distribution. We find that the ellipticity distributions of our LSBG samples (as well as the high surface brightness dwarf samples of \cite{kadofong2020}) are able to be well-reproduced by the relatively simple bivariate Gaussian model. We thus do not consider more complex functional forms in this work, though we have explored a model that includes size-shape covariance in \cite{kadofong2020} and the effect of a bimodal ground truth distribution on our bivariate and unimodal Gaussian modeling in an upcoming work (Kado-Fong et al., submitted).}
The observed count in bins of axis ratio is given as $n_i$, and $m_i$ is the
predicted count in the same range. We do not consider draws that lie outside of the observable boundaries ($\epsilon >0.7$ \rrr{for the HSC and DES samples} or \rrr{below the surface brightness limit}) -- that is, the imposed absence of high surface brightness or low $q$ sources does not
impact the likelihood computation.
 We adopt axis ratio bins of $dq=0.05$ and
mean surface brightness bins of $d\muavgmath{}=0.5$ for 
all inferences in this work\footnote{We previously found in \cite{kadofong2020}
that the choice of bin size does not strongly affect inference results}.

We use the Markov Chain Monte Carlo ensemble sampler implemented
in \textsf{emcee} \citep{foremanmackey2013} to sample efficiently from the 
posterior $\ln{}p(\vec\alpha, {\vec q_{\rm obs}},{\langle \vec \mu\rangle_{\rm R_{eff},obs}})$.
We implement a flat prior over the physical range of all our fitted parameters;
that is:
\begin{equation}
p(\rm \mba{})=
\begin{cases}
1 \quad{\rm if}\quad{} 0<\rm \mba{}<1 \\
0 \quad{\rm otherwise}
\end{cases}
\end{equation}
We additionally constrain $\rm \mca{}\leq\rm \mba{}$ to maintain the order of axes,
\begin{equation}
p(\rm \mca{})=
\begin{cases}
1 \quad{\rm if}\quad{} (0<\rm \mca{}<1)\wedge (\rm \mca{}\leq\rm \mba{}) \\
0 \quad{\rm otherwise},
\end{cases}
\end{equation}
When sampling from a given $\alpha$, we disregard ellipsoids where $C>B$.
We only require that the mean surface brightness is a positive value, i.e.
\begin{equation}
p(\rho_0)=
\begin{cases}
1 \quad{\rm if}\quad{} \mu_{\rho_0}>0 \\
0 \quad{\rm otherwise}.
\end{cases}
\end{equation}
Similarly, we require the standard deviations of
B/A and C/A are a positive value less than $\sigma=0.5$:
\begin{equation}
p(\sigma_X)=
\begin{cases}
1 \quad{\rm if}\quad{} 0<\sigma_X<0.5 \\
0 \quad{\rm otherwise}
\end{cases}
\end{equation}
for $X\in \{B,C\}$. We only require $\sigma_{\rho_0}$ to be positive:
\begin{equation}
p(\sigma_{\rho_0})=
\begin{cases}
1 \quad{\rm if}\quad{} 0<\sigma_{\rho_0} \\
0 \quad{\rm otherwise.}
\end{cases}
\end{equation}

\begin{deluxetable*}{lllllllll}
\tablecaption{\rrr{Median values and 95\% credible intervals} of inferred intrinsic shape parameters}
\tablehead{
\colhead{Associated Figure} &
\colhead{Sample} &      
\colhead{$\rm \mba{}$ (B/A)} & 
\colhead{$\sigma_{\mba{}}$} &
\colhead{$\rm \mca{}$ (C/A)} & 
\colhead{$\sigma_{\mca{}}$} &
\colhead{$\rm T_0$} &
\colhead{$\rho_0$} & 
\colhead{$\sigma_{\rho_0}$}
}
\startdata
\autoref{f:inferencecomparison} &    HSC-SSP LSBGs & $0.85^{+0.09}_{-0.07}$ & $0.07^{+0.07}_{-0.07}$ & $0.53^{+0.04}_{-0.04}$ & $0.13^{+0.04}_{-0.03}$ & $0.38^{+0.18}_{-0.24}$ &                     -- &                     -- \\
                                &        DES LSBGs & $0.88^{+0.01}_{-0.02}$ & $0.05^{+0.02}_{-0.01}$ & $0.55^{+0.01}_{-0.01}$ & $0.14^{+0.01}_{-0.01}$ & $0.34^{+0.07}_{-0.05}$ &                     -- &                     -- \\
                                & HI-selected UDGs & $0.80^{+0.10}_{-0.06}$ & $0.09^{+0.05}_{-0.08}$ & $0.50^{+0.08}_{-0.06}$ & $0.16^{+0.14}_{-0.06}$ & $0.50^{+0.14}_{-0.24}$ &                     -- &                     -- \\
                                &    red DES LSBGs & $0.91^{+0.03}_{-0.03}$ & $0.03^{+0.02}_{-0.03}$ & $0.55^{+0.02}_{-0.02}$ & $0.14^{+0.01}_{-0.02}$ & $0.25^{+0.08}_{-0.08}$ &                     -- &                     -- \\
                                &   blue DES LSBGs & $0.85^{+0.02}_{-0.03}$ & $0.05^{+0.02}_{-0.02}$ & $0.55^{+0.01}_{-0.01}$ & $0.14^{+0.01}_{-0.01}$ & $0.39^{+0.07}_{-0.06}$ &                     -- &                     -- \\
                                &        Coma UDGs & $0.90^{+0.07}_{-0.07}$ & $0.04^{+0.06}_{-0.04}$ & $0.54^{+0.06}_{-0.06}$ & $0.23^{+0.05}_{-0.05}$ & $0.26^{+0.18}_{-0.18}$ &                     -- &                     -- \\
          \autoref{f:sbcorrout} &        DES LSBGs & $0.87^{+0.02}_{-0.49}$ & $0.05^{+0.03}_{-0.04}$ & $0.48^{+0.04}_{-0.03}$ & $0.19^{+0.02}_{-0.12}$ & $0.31^{+0.10}_{-0.08}$ & $0.67^{+0.11}_{-0.50}$ & $0.40^{+0.05}_{-0.21}$ \\
                                &    red DES LSBGs & $0.83^{+0.06}_{-0.42}$ & $0.11^{+0.11}_{-0.10}$ & $0.53^{+0.45}_{-0.08}$ & $0.17^{+0.05}_{-0.06}$ & $0.43^{+0.26}_{-0.19}$ & $0.20^{+0.10}_{-0.08}$ & $0.50^{+0.08}_{-0.28}$ \\
                                &   blue DES LSBGs & $0.86^{+0.03}_{-0.05}$ & $0.05^{+0.03}_{-0.03}$ & $0.52^{+0.04}_{-0.04}$ & $0.18^{+0.03}_{-0.05}$ & $0.35^{+0.10}_{-0.09}$ & $0.60^{+0.16}_{-0.21}$ & $0.47^{+0.13}_{-0.08}$ \\
\enddata
\tablecomments{The top section shows the results of the ellipticity-only inference, while the bottom section shows the results of the surface brightness and ellipticity joint fit. In all cases, the first 250 steps of each walker are discarded. \rrr{For ease of comparison with literature results,
we also report $\rm T_0$, the median triaxiality ($\rm T = (1-(B/A)^2)/(1-(C/A)^2)$),
as calculated by
sampling the inferred shape distributions. }}
\end{deluxetable*}\label{t:inference}

We run all inferences for at least 500 steps over 32 walkers, and discard
the first 250 steps of each walker. We manually confirm that the chains
have converged. We also extend the mock recovery tests of \cite{kadofong2020}
to include this joint inference scheme -- we present the results of these tests
in \autoref{s:appendix}. We find that the joint inference reduces the 
inference precision for the spheroid and prolate populations, where
the surface brightness changes little with viewing angle, but increases the
inference precision for the disk population, where the surface brightness 
changes strongly with viewing angle.

\section{Results}\label{s:results}
We show the results of the intrinsic shape 
inference in \autoref{f:inferencecomparison}. 
From top left, we show the inference for the HSC-SSP LSBG sample of 
\cite{greco2018}, the full DES sample of \cite{tanoglidis2020}, the
\hi{}-selected sample originally constructed by \cite{leisman2017} and \cite{janowiecki2019}, 
the blue ($g-i<0.64$) DES LSBGs,
and the red ($g-i\geq0.64$) DES LSBGs. For the DES sample and subsamples, we 
show the inference results when surface brightness and ellipticity are 
jointly fit (unfilled turquoise contours) and when only ellipticity is considered
(filled orange histogram). We find that the joint fit does not converge
for the HSC-SSP and \hi{}-selected samples due to the relatively small sample sizes (a result that is expected based on the tests that we ran with mock galaxy populations, see \autoref{s:appendix} and \citealt{kadofong2020}); we thus
only show the results of the ellipticity-only inference.
In all cases, the joint maximum a posteriori (MAP) estimate and ellipticity-only
MAP estimates are within very good agreement, 
indicating that the observed surface brightness
cut does not significantly bias the observed ellipticity distribution. 
To rephrase this point, we
find that because the intrinsic shapes of the LSBGs are relatively round, 
the observed surface
brightness does not correlate strongly with the ellipticity.
We additionally show the distribution
of model surface brightnesses and ellipticities (along with their observational counterparts) for the DES samples in \autoref{f:sbcorrout}. We find that the surface brightness distribution of the red LSBGs is not as well-fit in our inference -- this likely indicates that the underlying surface brightness distribution is not well-described by a Gaussian. However, through tests with a uniform surface brightness distribution we find that our results are robust against a change in the assumed parametric form of the underlying surface brightness distribution. Furthermore, disky shapes are required to produce a strong correlation between surface brightness and ellipticity, and it is highly unlikely based on previous studies of dwarfs \citep{sanchezjanssen2010, burkert2017, rong2019, kadofong2020, carlsten2021} and higher mass galaxies \citep[see, e.g.][]{padilla2008} that the red LSBGs would be diskier than the blue LSBGs.
We present the maximum a posteriori estimates, along with associated uncertainties,
for all inference results in \autoref{t:inference}.

Though the numbers of the HSC-SSP and \hi{}-selected samples preclude a division as a function of galaxy color,
we can use these samples to confirm that our method returns consistent
results for both samples. For the HSC sample, 
this is a simple confidence check with a deeper sample -- 
assuming that the shapes measured in DES are not significantly impacted by the depth of the imaging, the samples should return a consistent result, as they are 
selected using similar methods. The concordance of \hi{}-selected
UDG intrinsic shape distribution is somewhat more intriguing, as the
sample is more luminous and selected to have significant stores of cold gas and
is likely a more luminous sample (see \autoref{f:massfig}) -- we defer a more complete discussion of this result to \autoref{s:discussion:observations}.

\subsection{Intrinsic Shape versus Color \rrr{and Environment}}
The correlated bimodality of morphology and color is a well-established facet of the
galaxy population for massive galaxies \citep[see, e.g.][]{padilla2008}
At dwarf masses, there is evidence that the bimodality persists in massive dwarfs \citep{kadofong2020}, with some evidence that the structural properties may begin to converge at lower masses \citep{carlsten2021}. Blue galaxies are typically 
also disk galaxies, while red galaxies are ellipticals. It is thus of interest
to ask whether the same bimodality is observed in the DES LSBG sample (the
HSC-SSP sample is not large enough to split into two samples).

As shown in \autoref{f:inferencecomparison}, 
we find no evidence for a significant difference in intrinsic shape 
distribution of the red and blue LSBGs of the DES sample. This result is not
unexpected; as shown in \autoref{f:sersiccomparison},
the \sersic{} index distribution of red and blue 
LSBGs are also remarkably similar for both the DES and HSC-SSP samples, as
well as the \hi{}-selected sample. This
is in strong contrast to more massive galaxies,
wherein the \sersic{} index distributions are markedly different between the red and blue galaxy populations.

\rrr{We also find that the three-dimensional shapes of the 
UDGs are largely unchanged as a function of environment, as shown by contrasting
the Coma UDGs (rightmost panel, second row in \autoref{f:comparison}) to the 
blue LSBGs and HI-selected galaxies, both of which are unlikely to be dominated
by dwarfs in cluster environments. Indeed, this result is in agreement with
previous measurements of cluster UDG shapes, as will be further discussed in
\autoref{s:discussion:observations}.}

\section{Discussion}\label{s:discussion}
Although the existence of LSBGs has been known observationally
for some time, the path through which this tail of the galaxy population is formed remains unclear. UDGs in the field have been proposed to form via a variety of processes, including that they may populate high spin halos \citep{dalcanton1997, amorisco2016, liao2019}, that they are formed via star formation feedback \citep{jiang2019, chan2018}, or that they are the end products of early mergers that cause star formation to migrate to large radii \citep{wright2020}. High density environments provide potential alternate pathways to UDG formation via environmental effects such as ram pressure stripping and tidal heating \citep{jiang2019, tremmel2020}. Here, we consider the implications of the inferred intrinsic shapes of our LSBG samples both in conjunction with previous observational works and in comparison to contemporary theoretical predictions.

\subsection{Comparison to Observations}\label{s:discussion:observations}
We first compare our inferred LSBG shape distribution to results from the 
literature. We probe both the potential influence of environment by 
comparing to the inferences of cluster and group UDG shapes from \cite{burkert2017}
and \cite{rong2019} and the potential influence of surface density by
comparing to the ``normal'' dwarf (meaning higher surface brightness)
spectroscopic sample of \cite{kadofong2020}, as shown in \autoref{f:comparison}.

We find remarkably good agreement between our results and those of \cite{rong2019},
who infer the shape distribution of UDGs in high density group and cluster 
environments. We also see good agreement with the results of \cite{burkert2017},
when it is considered that this work did not allow for triaxiality (i.e.
enforced B/A=C/A).
Though the environments of the DES and HSC-SSP samples are not
known on the level of individual systems, the spatial distribution of the red and blue
LSBGs of both samples suggest that the red LSBGs are more clustered
than the blue LSBGs, which have a nearly homogeneous distribution on the sky \citep{greco2018, tanoglidis2020}.
This indicates that red LSBGs are relatively more likely to live in 
high density environments, while blue LSBGs are relatively more likely to live
in the field. We can thus say that, on average, the blue LSBGs live in lower
density environments than the samples of \cite{burkert2017} and \cite{rong2019}.
Because these literature samples are focused on large groups and clusters, it
is also likely the case that the typical red LSBG lives in a lower density 
environment than \cite{burkert2017} and \cite{rong2019}, but in a higher density
environment than the typical blue LSBG. The concordance of shapes between all
these samples then indicates that the local environment does not play a key
role in determining the intrinsic shape of low surface brightness galaxies.

Next, we compare the results of the LSBG galaxy sample to those of the
spectroscopic ``normal'' dwarf sample of \cite{kadofong2020}. These dwarfs are 
drawn from the Sloan Digital Sky Survey (SDSS) and Galaxy And Mass Assembly (GAMA)
spectroscopic surveys. GAMA, the deeper of these surveys, has an effective
surface brightness limit of 
$\mu_{\rm r, eff} \sim 24$ \sbunits{} \citep{baldry2012},
meaning that the spectroscopic dwarf sample is nearly disjoint in
surface brightness with the LSBG samples. \cite{kadofong2020} finds 
evidence that the intrinsic shape of dwarf galaxies changes as a function of
mass -- in particular, that dwarfs at $M_\star \lesssim 10^{8.5}$ are relatively
more spheroidal than their higher mass analogs. To make a rough 
estimate of the stellar masses of our LSBG samples, we adopt a singular 
distance of 60 Mpc for the HSC-SSP sample (as informed by the cross-correlation
analysis in Greco et al., in prep.), and estimate the mass-to-light
ratio $(M/L)_i$ from the $(g-i)$ colors presented in \cite{greco2018} and the
color relations of \cite{bell2003}. We find that, for this rough approximation,
the mean stellar mass of the HSC-SSP sample is roughly $\log(M_\star/M_\odot)\sim 7.5$. 
Our HSC and DES LSBG samples are thus most similar in stellar mass to the lowest mass bin of 
$7.5<\log_{10}(\rm M_\star/M_\odot)<8.5$ in \cite{kadofong2020}. Indeed, we find
that these low mass, high surface brightness dwarfs are 
relatively oblate and spheroidal, similar to the LSBG samples
%This is in contrast
%to the higher mass dwarfs, which maintain well-formed, albeit thick, disks.
However, despite being likely more luminous than the HSC
and DES samples (see \autoref{f:massfig}), the \hi{}-selected sample is
also characterized by oblate spheroidal shapes. This is in contrast
to the observed mass evolution of normal dwarfs, wherein
 the higher mass dwarfs maintain well-formed, albeit thick, disks.
We thus suggest that the LSBGs may be thicker (in C/A) than the equivalent HSB
galaxy sample -- this effect is most distinct when considering that the more luminous
\hi{}-selected sample is also characterized by oblate spheroidal shapes. However, the incompleteness of the HSB sample at low
$(M_\star\lesssim 10^8 M_\odot)$ masses and the uncertainties in the
stellar mass distribution of the HSC and DES samples suggest that \rrr{more complete samples of HSB dwarfs, along with more distance determinations for LSBG dwarfs, are needed
to secure this result}.

Taken all together, the results presented in this work in conjunction with 
literature results from group/cluster UDGs and normal dwarfs shed new light 
onto the structure formation of LSBGs. 
%In particular, 
%the concordance of 
%shapes between LSBGs and similarly massive ``normal'' dwarfs
%implies that the formation mechanism of LSBGs does not radically change the
%shape distribution seen at higher surface brightnesses. 
First, the concordance of
shapes between LSBGs as a function of environment implies that the 
formation mechanism of cluster and 
field LSBGs does not produce drastically
different intrinsic shapes. Second, we suggest that LSBGs and UDGs may be
rounder than their HSB counterparts, and
possibly unable to support well-formed stellar disks where HSB dwarfs succeed
in maintaining them.

\subsection{Comparison to Simulations}
Having contextualized our results with previous observational results from the
literature, we now compare our findings to predictions from the UDG
populations produced in cosmological simulations. There are several paths to
UDG formation proposed by various simulations; we focus here on the Auriga 
simulation \citep{liao2019}, the NIHAO simulation \citep{jiang2019}, and
the Romulus simulations \citep{tremmel2020, wright2020}.

The UDG formation path of the simulations differ significantly,
and those differences manifest in the predicted intrinsic shape distribution of
each simulation. The Auriga field UDGs form in high spin halos (see also
\citealt{dalcanton1997,amorisco2016} who also form UDGs in the high spin tail the
halo population), while satellite UDGs form via a mixture of tidal effects and field UDG
capture by massive halos. This results in a pronounced difference in the intrinsic
shape distributions of the red and blue UDGs, wherein the blue Auriga UDGs are 
thick disks (turquoise point, right panel of \autoref{f:comparison}) and
the red Auriga UDGs are spheroidal (orange point). This shape contrast is in 
disagreement with our results, which do not point to a significant bimodality in
shape as a function of color or a significant population of disky UDGs
at any mass. Furthermore, our results indicate that the \hi{}-selected LSBG sample is puffier (higher C/A) than the HSB sample of \cite{kadofong2020} at similar stellar masses -- this
finding is also at odds with the theory that LSBGs form in the high angular momentum tail of the halo distribution function.

The NIHAO simulations, meanwhile, predict that UDGs are formed in the field
via supernovae feedback. These UDGs are characterized by particularly bursty
star formation histories relative to the more compact galaxies in NIHAO 
\citep{dicintio2017}.
Satellite UDGs are formed from infalling field UDGs 
and created from tidal effects. Though \cite{jiang2019} do not compute the
intrinsic shape distributions of satellite and field (or red and blue) UDGs
separately, they do report the overall mean intrinsic principal axis ratios.
We find that the NIHAO results are significantly more triaxial than our
observed results. This excessive triaxiality is also seen in a subset of the
FIRE galaxies which are also characterized by particularly bursty
star formation histories \citep[][Kado-Fong et al., in prep]{chan2018}.
\rrr{\cite{cardonabarrero2020} further analyze the shape distribution of the NIHAO UDGs
as a function of their stellar kinematics. They find in particular that the 
rotationally supported UDGs are characterized by more oblate (B/A$_{\rm rotation} >$ B/A$_{\rm dispersion}$) and flatter disks. We find that these rotationally supported UDGs are,
on average, flatter than our observed LSBG and UDG samples, similar to those
found in the Auriga simulations. }

The formation mechanisms for UDGs in the Romulus simulations stand in contrast to those presented in NIHAO and Auriga. Since Romulus cannot resolve high-density star formation and thus large outflows in dwarfs, feedback cannot drive their formation. However, 
upcoming comparison work between the Romulus simulations and the Marvel-ous Dwarfs, a zoom suite that succeeds in forming cored dwarf profiles via feedback with a force resolution of 60 pc and a DM [stellar] mass resolution of 6650 $\rm M_\odot$ [420 $\rm M_\odot$] \citep{munshi2021}, does indicate that the UDG shapes of Romulus and Marvel-ous are broadly consistent (Munshi et al., in prep.). Thus, we do not expect that the intrinsic shapes of the LSBGs in Romulus are simply an effect of resolution. The Romulus simulations furthermore do not find halo spin as a primary mechanism for UDG formation, as is the case in Auriga.  

Both \citet{wright2020} and \cite{tremmel2020} explore alternative formation mechanisms in isolation and in a cluster environment. In particular, \cite{tremmel2020} suggest that cluster UDGs are formed primarily through the dual effects of passive fading following quenching via ram pressure after early cluster infall and size evolution in the cluster environment. In the field, \cite{wright2020} find that Romulus UDGs are formed via early major mergers that redistribute star formation to larger radii. Van Nest, J. et al., in prep will explore the shape evolution of UDGs in comparison to dwarfs in Romulus. 
We select LSBGs from Romulus25 \citep{tremmel2017} and RomulusC \citep{tremmel2019} on their central surface brightness and effective radius in a method similar to that in \citet{tremmel2020, wright2020}. To best match the properties of the LSBG sample in this work,
we impose a size cut of \reff{}$>1$ kpc and a surface brightness cut of $\muavgmath>24.3$ \sbunits{}.
They are divided into blue and red populations based on their $g-i$ color using the same
dividing value of $(g-i)=0.64$ as the observational sample. The values shown in \autoref{f:comparison} correspond to the median of each population with the error bars representing the \rrr{16}th to \rrr{84}th percentile ranges.  
Although
the Romulus LSBGs show slightly more intrinsic shape evolution as a function of their $(g-i)$ color, \rrr{both the red and blue simulated LSBG populations are in reasonable agreement with our observed samples.}
We expect that the color-shape evolution seen in Romulus is partially driven by resolution effects: dwarfs are overquenched in the cluster environment of RomulusC \citep{tremmel2017}, leading to overly red colors and an over-representation of cluster LSBGs at the red end of the color distribution. We thus conclude that the Romulus LSBGs are in the best agreement with the inferred observational
shapes out of the simulation results considered here. The intrinsic shapes of the
blue LSBGs, \rrr{whose formation path is not tied to being in a high-density environment,} are in particularly good agreement with the inferred shapes of the observational
samples, \rrr{which also sample a wide range of environments}.

\iffalse
The Romulus simulation suite produces UDGs by yet another pathway. The UDGs
in Romulus25 are formed via early major mergers that cause star formation to
migrate outwards \citep{wright2020}. The Romulus UDGs are characterized as 
oblate spheroids, with principal axis ratios in good agreement with our 
observational results. 
\fi

\section{Conclusions}
In this work we have presented the first three-dimensional shape inference 
of the wide-field samples of LSBGs selected in DES, HSC-SSP, and ALFALFA. 
We find that all three samples
are well-characterized by oblate spheroids, with minor principal axis ratios
(C/A) significantly higher than the thick disks observed in high mass, \rrr{high surface brightness}
dwarfs. We also find no significant difference in the shape distribution
of red and blue LSBGs, an inference bolstered by the analogous concordance 
in the distribution of \sersic{} indices over red and blue LSBGs. 

Our inferred shape distributions are in good agreement with the shape 
distributions inferred for cluster UDGs \citep{burkert2017, rong2019}. 
These results suggest that the intrinsic shapes of LSBGs are not greatly 
affected by their environments. We also find some evidence that LSBGs are 
unable to maintain stellar disks at stellar masses where normal, high surface
brightness dwarfs regularly maintain thick disks. Intriguingly, we do note that there is evidence that these UDGs are still able to support gaseous disks \citep{mancerapina2019b, mancerapina2020, gault2021}. However, more work is
needed to address both the mass incompleteness of HSB dwarf samples below
$M_\star\sim10^8 M_\odot$ and the large uncertainties in
the distance estimate of the LSBG samples.
%, and that their formation channel does not 
%cause them to diverge significantly in form from their higher surface
%brightness analogs.

The different formation mechanisms proposed by cosmological simulations 
for UDGs and LSBGs manifest strongly in their intrinsic shape 
distributions. We find that our results are in some conflict with simulated
UDGs that form in high spin halos and those that are puffed up through
vigorous star formation feedback. However, we do find that our observed
intrinsic shape distribution is in good agreement with simulated UDGs from
the Romulus simulation suite, wherein field UDGs are formed
by early major mergers that cause star formation to migrate outwards \citep{wright2020} and cluster UDGs are formed via
environmental quenching and tidal heating \citep{tremmel2020}. 

These results show that intrinsic shape distributions are able to 
provide promising constraints on the formation path of the extreme
low surface brightness end of the dwarf galaxy population. The results in this work suggest that the intrinsic shapes of the LSB and HSB galaxy populations may begin to converge at low masses. However, future 
observational efforts to contextualize the LSBG population face two 
substantial technical challenges: linking LSBG and HSB samples in order to
systematically characterize dwarf properties across the spectrum of
surface brightnesses, and establishing distance measures to LSBGs in the field.
Upcoming survey instruments (e.g. Rubin/LSST, PFS, DESI, WALLABY, SKA,) will provide
the necessary sensitivity and survey power to make significant strides
in the former.
Work on the latter is an ongoing area of development (Greco, J. et al., in prep.), and
will allow analyses such as the one presented in this work to provide a truly
comprehensive view of the diversity in dwarf structure.

\acknowledgements
We thank Anna Wright, Alyson Brooks, \rrr{and Arianna di Cintio} for insightful
conversations which have improved the quality of 
this manuscript. \rrr{We also thank the anonymous referee for a helpful review of this work that helped to strengthen this analysis.} We acknowledge the support of the National Astronomy Consortium (NAC) for supporting the summer research of MM at Princeton University. J.P.G. is supported by an NSF Astronomy and Astrophysics Postdoctoral Fellowship under award AST-1801921. EAKA is supported by the WISE research programme, which is financed by the Dutch Research Council (NWO)

The Legacy Surveys consist of three individual and complementary projects: the Dark Energy Camera Legacy Survey (DECaLS; Proposal ID 2014B-0404; PIs: David Schlegel and Arjun Dey), the Beijing-Arizona Sky Survey (BASS; NOAO Prop. ID 2015A-0801; PIs: Zhou Xu and Xiaohui Fan), and the Mayall z-band Legacy Survey (MzLS; Prop. ID 2016A-0453; PI: Arjun Dey). DECaLS, BASS and MzLS together include data obtained, respectively, at the Blanco telescope, Cerro Tololo Inter-American Observatory, NSF's NOIRLab; the Bok telescope, Steward Observatory, University of Arizona; and the Mayall telescope, Kitt Peak National Observatory, NOIRLab. The Legacy Surveys project is honored to be permitted to conduct astronomical research on Iolkam Duag (Kitt Peak), a mountain with particular significance to the Tohono Oodham Nation.

NOIRLab is operated by the Association of Universities for Research in Astronomy (AURA) under a cooperative agreement with the National Science Foundation.

This project used data obtained with the Dark Energy Camera (DECam), which was constructed by the Dark Energy Survey (DES) collaboration. Funding for the DES Projects has been provided by the U.S. Department of Energy, the U.S. National Science Foundation, the Ministry of Science and Education of Spain, the Science and Technology Facilities Council of the United Kingdom, the Higher Education Funding Council for England, the National Center for Supercomputing Applications at the University of Illinois at Urbana-Champaign, the Kavli Institute of Cosmological Physics at the University of Chicago, Center for Cosmology and Astro-Particle Physics at the Ohio State University, the Mitchell Institute for Fundamental Physics and Astronomy at Texas A\&M University, Financiadora de Estudos e Projetos, Fundacao Carlos Chagas Filho de Amparo, Financiadora de Estudos e Projetos, Fundacao Carlos Chagas Filho de Amparo a Pesquisa do Estado do Rio de Janeiro, Conselho Nacional de Desenvolvimento Cientifico e Tecnologico and the Ministerio da Ciencia, Tecnologia e Inovacao, the Deutsche Forschungsgemeinschaft and the Collaborating Institutions in the Dark Energy Survey. The Collaborating Institutions are Argonne National Laboratory, the University of California at Santa Cruz, the University of Cambridge, Centro de Investigaciones Energeticas, Medioambientales y Tecnologicas-Madrid, the University of Chicago, University College London, the DES-Brazil Consortium, the University of Edinburgh, the Eidgenossische Technische Hochschule (ETH) Zurich, Fermi National Accelerator Laboratory, the University of Illinois at Urbana-Champaign, the Institut de Ciencies de l'Espai (IEEC/CSIC), the Institut de Fisica d'Altes Energies, Lawrence Berkeley National Laboratory, the Ludwig Maximilians Universitat Munchen and the associated Excellence Cluster Universe, the University of Michigan, NSF's NOIRLab, the University of Nottingham, the Ohio State University, the University of Pennsylvania, the University of Portsmouth, SLAC National Accelerator Laboratory, Stanford University, the University of Sussex, and Texas A\&M University.

BASS is a key project of the Telescope Access Program (TAP), which has been funded by the National Astronomical Observatories of China, the Chinese Academy of Sciences (the Strategic Priority Research Program The Emergence of Cosmological Structures Grant XDB09000000), and the Special Fund for Astronomy from the Ministry of Finance. The BASS is also supported by the External Cooperation Program of Chinese Academy of Sciences (Grant 114A11KYSB20160057), and Chinese National Natural Science Foundation (Grant 11433005).

The Legacy Survey team makes use of data products from the Near-Earth Object Wide-field Infrared Survey Explorer (NEOWISE), which is a project of the Jet Propulsion Laboratory/California Institute of Technology. NEOWISE is funded by the National Aeronautics and Space Administration.

The Legacy Surveys imaging of the DESI footprint is supported by the Director, Office of Science, Office of High Energy Physics of the U.S. Department of Energy under Contract No. DE-AC02-05CH1123, by the National Energy Research Scientific Computing Center, a DOE Office of Science User Facility under the same contract; and by the U.S. National Science Foundation, Division of Astronomical Sciences under Contract No. AST-0950945 to NOAO.

\bibliography{LSBGintrinsic.bib}

\appendix

\section{Mock Recovery Tests}\label{s:appendix}

\begin{figure*}[htb]
\centering     
\includegraphics[width=\linewidth]{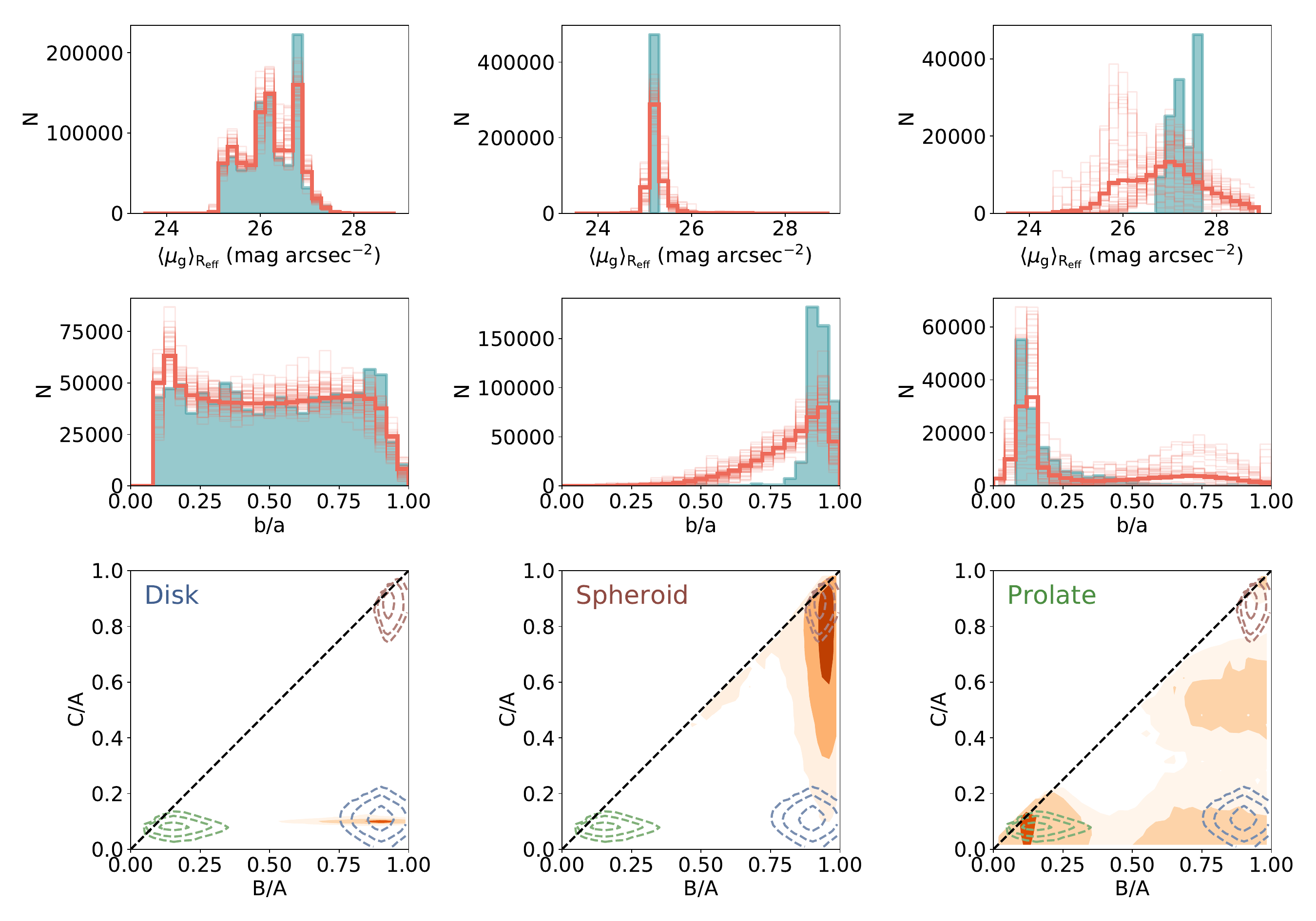}
%\vspace{-25pt}
\caption{ 
    Recovery results for joint inference of the three extreme mock populations
    presented in \cite{kadofong2020}. Each column shows a
    different mock population; from left to right, we show the
    disk, spheroid, and prolate mock populations. \textit{Top row:} The
    surface brightness distribution for each population. The thick orange curve
    show our posterior sample (thin orange curves show individual pulls from
    the posterior). The solid \rrr{teal} histograms show the observed population.
    \textit{Middle row:} the same for the ellipticity distributions.
    \textit{Bottom row:} we show the intrinsic axis ratio distribution of
    the posterior sample as filled orange contours. The analogous results
    from ellipticity-only inferences are shown by the dashed contours.
    }
\label{f:mockrecovery}
\end{figure*}

We extend the mock recovery tests of \cite{kadofong2020} to include 
the relation between surface brightness and intrinsic shape in our mock
recovery tests. We implement this relation by computing the expected
surface brightness of each synthetic population given the intrinsic shape
and viewing angle of each mock galaxy. We do not impose surface brightness
limits for these mock tests. \rrr{Because 
the mock galaxies projected structural parameters are not remeasured from mock HSC imaging, the 
absolute value of the surface brightnesses in this test are not consequential to the results. Rather, 
the test is designed to evaluate our ability to recover the shape of the surface brightness distribution.}
In \autoref{f:mockrecovery} we
show the inferred intrinsic shape distribution of the posterior sample,
as well as the distribution of the model population in average surface
brightness (top row) and ellipticity (middle row). We find that
the joint inference, on average, makes the inference more uncertain. 
However, the inference actually becomes more precise for the disky mock
population. This can be understood by considering that the correlation
between surface brightness, ellipticity, and viewing angle is maximized
for a disky object. \rrr{The inference becomes more 
precise due to the additional information provided by this covariance. For the spheroidal and
prolate populations, however, the surface brightness distributions are very narrow. Thus, although
two extra parameters must now be fitted, the surface brightness provides relatively
little extra information.}

\end{document}